\documentclass[prd,aps,preprint,amsmath,nofootinbib,amssymb,eqsecnum,showkeys,tightenlines]{revtex4-1}

\usepackage{amsmath, latexsym, amssymb, hyperref, graphicx, color, multirow, makecell, diagbox}
\usepackage[table]{xcolor}
\usepackage{tabularx}
\usepackage[T1]{fontenc}
\usepackage[capitalize]{cleveref}
\definecolor{nicered}{rgb}{.7,.1,.1}
\definecolor{nicegreen}{rgb}{.1,.5,.1}
\definecolor{darkblue}{rgb}{0,0,.5}
\hypersetup{colorlinks, citecolor=nicegreen,linkcolor=nicered, urlcolor=darkblue}
\usepackage{multirow}
\usepackage{slashed}
\numberwithin{equation}{section}
\usepackage{verbatim}
\usepackage{graphicx}
\usepackage{CJKutf8}
\graphicspath{{picture/}, {picture/D04/}, {picture/S04/}}

\begin{document}
\preprint{}

\title{Detecting anomaly in vector boson scattering
}

\author{Jinmian Li(\begin{CJK}{UTF8}{gbsn}李金勉\end{CJK})$^{1}$}
\email{jmli@scu.edu.cn}

\author{Shuo Yang(\begin{CJK}{UTF8}{gbsn}杨硕\end{CJK})$^{2,3}$}
\email{shuoyanglnnu@163.com}

\author{Rao Zhang(\begin{CJK}{UTF8}{gbsn}张饶\end{CJK})$^{1}$}
\email{zhangrao@stu.scu.edu.cn}

\affiliation{$^1$ College of Physics, Sichuan University, Chengdu 610065, China}
\affiliation{$^2$ Department of Physics, Liaoning Normal University, Dalian 116029, China}
\affiliation{$^3$ Department of Physics, Dalian University, Dalian 116622, China}

\begin{abstract}
Measuring the vector boson scattering (VBS) precisely is an important step towards understanding the electroweak symmetry breaking of the standard model (SM) and detecting new physics beyond the SM.
We propose a neural network which compress the features of the VBS into three dimensional latent space.
The consistency of the SM prediction and the experimental data is tested by the binned log-likelihood analysis in the latent space.
We will show that the network is capable of distinguish different polarization modes of $WWjj$ production in both dileptonic channel and semi-leptonic channel.
The method is also applied to constrain the effective field theory and two Higgs Doublet Model. The results demonstrate that the method is sensitive to generic new physics contributing to the VBS.

\end{abstract}


\maketitle

\section{Introduction}\label{sec:intro}

Vector Boson Scattering (VBS) represents sensitive probe of both the Standard Model (SM) electroweak symmetry breaking (EWSB) and new physics Beyond-the-SM (BSM)~\cite{Rauch:2016pai,Green:2016trm}.
If the couplings of the Higgs boson to vector bosons deviate from the SM prediction, the cross sections of VBS processes will increase with center-of-mass energy up to the scale of new physics.
In addition, many BSM models predict extended Higgs sector. The contribution from new resonances can also increase the VBS cross section in certain phase space.

Measuring the VBS processes at hadron collider is experimentally challenging due to their low signal yields and complex final states. The LHC experiments have built comprehensive searches for the VBS processes~\cite{Alessandro:2018khj,Baglio:2020bnc,Gallinaro:2020cte}.
The same-sign $WW$ production with leptonic decay has the largest signal-to-background ratio among VBS processes. This channel was the first VBS process that has been observed during the run 1 of the LHC~\cite{Aad:2014zda,Khachatryan:2014sta} and has been confirmed by the measurements at the LHC run II~\cite{Aaboud:2019nmv,Sirunyan:2017ret}.
The ATLAS and CMS Collaborations have also performed the measurements for other VBS channels, such as fully leptonic $ZZ$~\cite{Sirunyan:2017fvv,Aad:2020zbq},  fully leptonic $WZ$~\cite{Aaboud:2018ddq, Sirunyan:2019ksz} and semi-leptonic $WV$ or $ZV$ with the $V$ decaying hadronically~\cite{Aad:2019xxo,Sirunyan:2019der}.
New physics contributions to the VBS channels are usually parameterized by effective field theory (EFT) operators. Precision measurement of the VBS channels can be recast as constraints on the coefficient of the operators~\cite{Fabbrichesi:2015hsa,Liu:2018pkg,Stolarski:2020qim}.

Understanding the polarization of the gauge bosons is an important step after the measurements of the VBS processes.
Vector bosons are unstable and can only be observed through their decay products. This lead to the interference among different polarizations, which cancels exactly only when the azimuthal angles of the decay products are integrated over. Even though selection cuts in analyses render the incompleteness of the cancellation, it is still possible to extract polarization fractions by fitting data with Monte Carlo simulated templates.
There are studies aiming to determine the polarization of gauge bosons in the $W^\pm W^\mp$ channel~\cite{Han:2009em,Ballestrero:2017bxn}, in fully leptonic $W^\pm W^\pm$ channel~\cite{Ballestrero:2020qgv}, in fully leptonic WZ/ZZ channels~\cite{Ballestrero:2019qoy}, in the SM Higgs decay~\cite{Maina:2020rgd} and in generic processes with boosted hadronically decaying $W$ boson~\cite{De:2020iwq}.
Various kinematic observables have been proposed in these works to discriminate the longitudinal and transverse polarized gauge boson.
Several recent studies have shown that deep neural network with input of final states momenta can be used for regression of the lepton angle in the gauge boson rest frame~\cite{Searcy:2015apa,Grossi:2020orx} and classification of events from different polarizations~\cite{Lee:2018xtt,Lee:2019nhm}.

Autoencoders have been widely used in model-agnostic searches at colliders, dubbed as anomaly detection or novelty detection. The main function of the autoencoder is that it learns to map an input to a latent compressed representation and then back to itself. The autoencoder which is trained on known SM processes could be able to identify the BSM events as anomalies~\cite{Cerri:2018anq,Collins:2018epr,Collins:2019jip, Blance:2019ibf,Andreassen:2020nkr,Nachman:2020lpy,Collins:2019jip,Farina:2018fyg,Roy:2019jae}.
In other cases, when the anomaly can not be detected on a single event, density-based novelty evaluators~\cite{DAgnolo:2018cun,DeSimone:2018efk,Hajer:2018kqm} are proposed to detect discrepancies between two datasets in the latent space.
Since the VBS processes are the perfect window to access any new physics related with EWSB, we can adopt autoencoders to detect possible new physics contributions to the process.

In this work, focusing on the fully leptonic and semi-leptonic channels of the $W^\pm W^\mp$+jets process, we propose a neural network based on the Transformer architecture~\cite{vaswani2017attention} to learn the features of the VBS process. Those features are not only useful in separating the VBS process from the SM backgrounds but also capable of discriminating different polarizations of the $W$ bosons in the VBS process.
An autoencoder is trained on the features to reduce the dimensionality so that only the most relevant features are kept. Eventually, we perform binned log-likelihood test in the latent space to find out whether the distributions of the feature is coincide with the SM prediction. The EFT and Two Higgs Doublet Model (2HDM) are considered as examples to demonstrate that this method is able to test a wide class of BSM physics.

The paper is organized as follows. The analysis framework is introduced in Sec.~\ref{sec:framework}, including the event generation, architecture of neural network and binned log-likelihood analysis. Discrimination of different polarization modes of the $WWjj$ production is discussed in Sec.~\ref{sec:pol}. In Sec.~\ref{sec:eft} and Sec.~\ref{sec:2hdm}, we consider the applications of our method to effective field theory and two Higgs Doublet Model, respectively.
Our conclusions are presented in Sec.~\ref{sec:conclude}.


\section{Analysis framework}\label{sec:framework}

\subsection{Event generation for signals and backgrounds}\label{sec:events}

The signal and background events in our study is generated with the \textsc{MadGraph5\_{}aMC@NLO} \cite{Alwall:2014hca} framework, in which the \textsc{Madspin} is used for the decays of heavy SM particles (top quark, W/Z boson), and \textsc{Pythia 8.2}~\cite{Sjostrand:2007gs}  is used for parton shower, hadronization and decay of hadrons.
The latest version of \textsc{MG5} is capable of handling polarized parton scattering~\cite{BuarqueFranzosi:2019boy}. This function is adopted to simulate the events of the VBS processes with fixed vector boson polarization in the final state.
The detector effects are simulated by \textsc{Delphes 3} with ATLAS configuration card, where $b$-tagging efficiency is set to 70\%, and mistagging rates for the charm- and light-flavor jets are 0.15 and 0.008, respectively~\cite{ATLAS:2016gsw}.
The clustering of final state particles into jets are implemented by \textsc{FastJet}~\cite{Cacciari:2011ma} using the anti-$k_T$ algorithm with cone size parameter $R=0.4$.

All of the diagrams at $\alpha_{\text{EW}}^4$ ($\alpha_{\text{EW}}$ is the electroweak coupling constant) are included in simulating the VBS process (referred as EW production hereafter), such as $\gamma \gamma \to WW$, processes with final state vector boson radiated from quark directly, and the significant interferences among diagrams.
There are also mixed electroweak-QCD diboson productions at $\mathcal{O}(\alpha_s^2 \alpha^2_{\text{EW}})$, where $\alpha_s$ is the strong coupling constant. In the SM, the interference between the electroweak and mixed EW-QCD production is found to be small~\cite{Biedermann:2017bss,Ballestrero:2017bxn,Campanario:2020xaf}.
In simulating the polarized processes, the definition of the polarization is frame-dependent. We take the partonic center of mass frame as the reference frame in this work, i.e. the rest frame defined by the two initial parton in the $q q'\to W^+ W^- j j$ process~\footnote{One could also use the rest frame of $W^+ W^-$ system as the reference frame, in which the fraction of longitudinal polarized $W$ boson is slightly higher~\cite{BuarqueFranzosi:2019boy}. }.

We will study both the dileptonic channel and semi-leptonic channel of the EW $W^\pm W^\mp jj$ production. So that at least one of the $W$ bosons should be decaying leptonically (denoted by $W_\ell W jj^{\text{EW}}$).
The dominant backgrounds are QCD production of $t\bar{t}$ process, single top production, mixed EW-QCD production of $WW/WZ$ and the EW production of $WZ$.
Since the fully hadronic final states are not relevant in our analysis, the following requirements are applied in generating the background  events: (1) at least one of the top decays leptonically in the $t\bar{t}$ process (denoted by $t t_\ell$); (2) either $W$ or top quark decays leptonically in the $tW$ process (denoted by $tW_\ell/t_\ell W$) ; (3) at least one of  the $W$ boson decays leptonically in the mixed electroweak-QCD $WWjj$ process (denoted by $W_\ell Wjj^{\text{QCD}}$); (4) the $W$ boson decays leptonically in the mixed electroweak-QCD $WZjj$ process (denoted by $W_\ell Zj j^{\text{QCD}}$) and in the EW $WZjj$ process (denoted by $W_\ell Z j j^{\text{EW}}$).
In all of those cases, the transverse momenta of final state jets should be greater than 20 GeV.
We will use the measured inclusive cross sections at the LHC for $t\bar{t}$~\cite{CMS:2016rtp} and $tW$~\cite{Sirunyan:2018lcp} processes,
and use the leading order cross sections which are calculated by \textsc{MadGraph5\_{}aMC@NLO} for diboson processes.
The fiducial cross sections at 13 TeV LHC are provided in the second column of Tab.~\ref{tab:xsec}.

\begin{table}[h!]
\begin{center}
\begin{tabular}{|c||c|c|c|} \hline
  & $\sigma^{\text{fid}}$ [pb] & $\sigma^{\ell \ell}$ [fb] & $\sigma^{\ell j}$ [fb] \\ \hline
  $t t_\ell$                          &     210.3 & 139.8 & 3007.6 \\ \hline
  $t W_\ell$/$t_\ell W$     &     15.9  &  11.6  &  224.6 \\ \hline
  $W_\ell W j j^{\rm QCD}$ &    4.68  &  14.7 &  340.5 \\ \hline
  $W_\ell Z j j^{\rm QCD}$  &    2.20  &   4.49 & 165.7\\ \hline
    $W_\ell Z j j^{\rm EW}$  &  0.487 &  3.68  & 22.2\\ \hline
    $W_\ell  Wj j^{\rm EW}$ &   0.738 &  4.36  & 37.3 \\ \hline
\end{tabular}
\caption{\label{tab:xsec}  The production cross sections of signal and background processes before and after pre-selections.  }
\end{center}
\end{table}

The events are divided into two classes with the following preselections~\cite{Alessandro:2018khj}:
\begin{itemize}
\item \textbf{Di-Lepton:}  exactly two opposite sign leptons with $p_T(\ell) >20~\text{GeV}, ~ |\eta(l)|<2.5$; at least two jets with $ p_T(j) >20~\text{GeV}, ~ |\eta(j)|<4.5$; the two jets with leading $p_T$ should give large invariant mass ($m_{jj}>500$ GeV) and have large pseudorapidity separation ($|\Delta \eta|_{jj}>3.6$); no $b$-tagged jet in the final state.
\item \textbf{Semi-Lepton:} exactly one charged lepton with $p_T(\ell) >20~\text{GeV}, ~ |\eta(l)|<2.5$; at least four jets with  $ p_T(j) >20~\text{GeV}, ~ |\eta(j)|<4.5$; the pair of jets with the largest invariant mass ($m_{jj}>500$ GeV) that also satisfies $|\Delta \eta|_{jj}>3.6$ is taken as the forward-backward jet pair; (4) among the remaining jets, the jet pair with invariant mass closest to the $W$ boson mass is regarded as the jet pair from $W$ decay.
\end{itemize}
The cross sections for signal and backgrounds after the Di-Lepton and Semi-Lepton selections are provided in the third and fourth columns of the Tab.~\ref{tab:xsec}, respectively.
We can find that the $t\bar{t}$ process is the most important background in both channels, the cross section of which is $\sim \mathcal{O}(100)$ times larger than that of the VBS process.

The preselected events are fed into the network for learning the features.
The deep learning is known to be able to transform lower level inputs into discriminative outputs.
So we represent each event by a set of four-momenta~\footnote{We use the $(p_x, p_y, p_z, E)$, although sometimes $(p_T, \eta, \phi, m)$ is used. } and their identities (the lepton charge is implied).
Different networks will be adopted for dileptonic channel and semi-leptonic channel. The input for the network of dileptonic channel consists of momenta of two leptons, forward and backward jets, sum of all detected particles and sum of jets that are not assigned as forward-backward jets.
And the input for the network of semi-leptonic channel consists of momenta of the lepton, forward and backward jets, two jets from $W$ decay, sum of all detected particles and sum of remaining jets~\footnote{Jets that are not assigned as forward-backward jets and jets from $W$ boson decay.}.
In short, there are six/seven momenta with identities for the input of dileptonic/semi-leptonic channel.

\subsection{Architecture of neural network}
A simple fully connected neural network can extract the features of the input data, but there are a lot of redundant connections, which will make the extraction efficiency low and prone to overfitting.
These problems can be alleviated by including the attention mechanism.
As proposed in Ref.~\cite{vaswani2017attention}, the Transformer with multi-head self-attention mechanism provides a variety of different attentions and improved the learning ability, thus can be used to  effectively extract the internal connections of features.

\begin{figure}[htbp]
\includegraphics[width=0.9\textwidth]{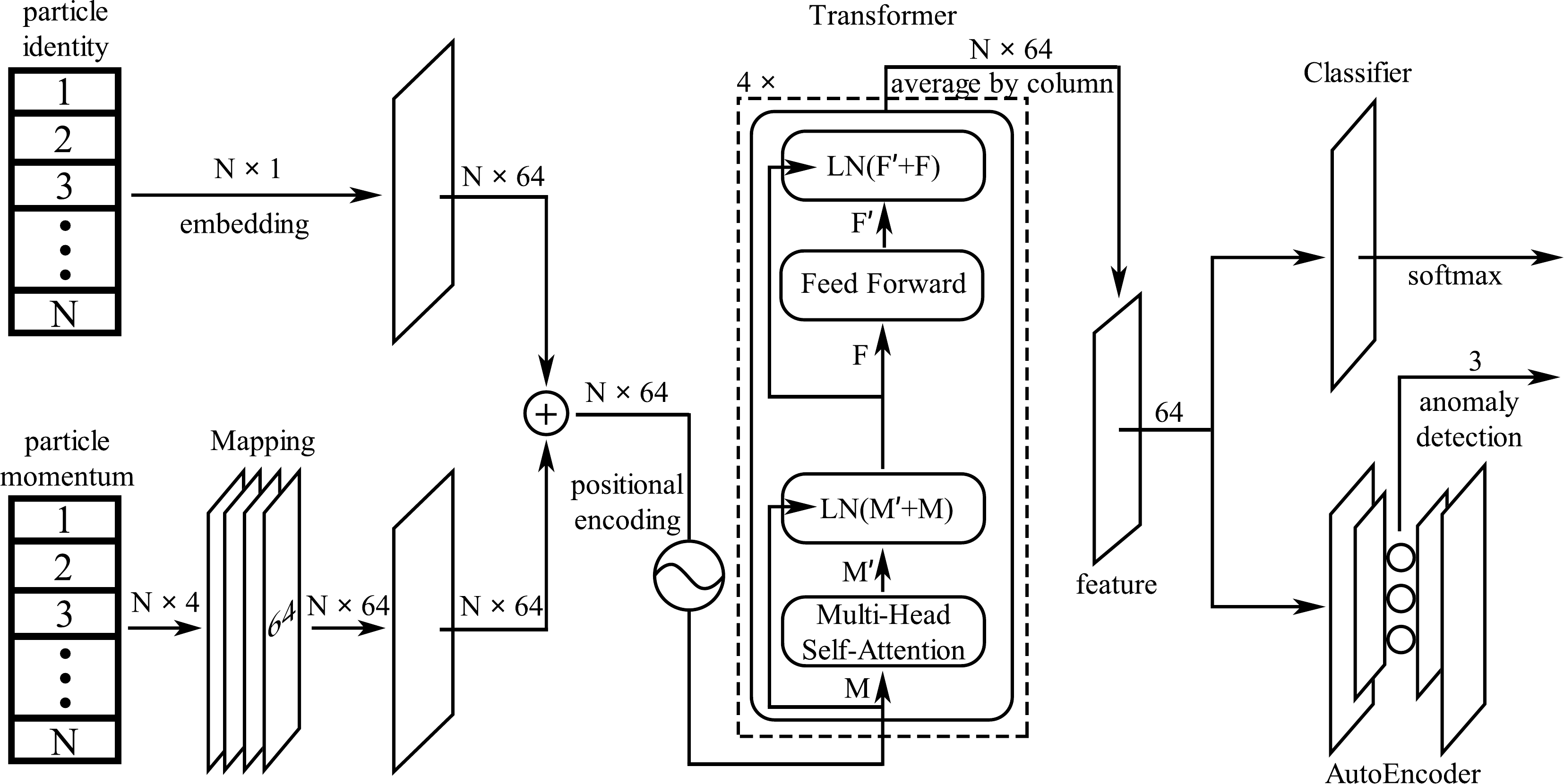}
\caption{Architecture of neural network. \label{fig:network}}
\end{figure}

The architecture of our neural network is illustrated in Fig.~\ref{fig:network}.
The input consists of identities and four-momenta of $N$ particles ($N=6/7$ for the dileptonic/semi-leptonic channel).
The original particles momentum ($p^{\mu}$) is normalized according to
\begin{align}
\hat{p}^{\mu}_{i} = \frac{p^{\mu}_{i} - \bar{p}^\mu} {\sigma_{p^{\mu}}},
\end{align}
where the index $i$ runs over $N$ particles in an event.
The mean $\bar{p}^\mu$ and standard deviation $\sigma_{p^{\mu}}$ are calculated on the particles from the full set of the training sample.
Then, we embed the particles identities of each event into a uniform distribution ($N \times 64$), and map the normalized four momenta to a matrix ($N\times64$) through a Mapping network. The Mapping network is a fully connected neural network with 4 hidden layers (each layer contains 64 neurons).
The summation of those two components (encode the types of particles into the four momenta, denoted by $M_{N\times 64}$) are fed into the Transformer.
The Transformer contains four copies of encoder layers. Each encoder consists of a self-attention layer and a feed forward neural network followed by normalization layers.
In particular, the self-attention layer map the $M_{N\times 64}$ into $M'_{N\times 64}$

\begin{align}
M'_{N\times 64} = [\text{Softmax}(\frac{W^Q_1 (W^K_1)^T}{8 }) W^V_1, \cdots ,\text{Softmax}(\frac{W^{Q}_4 (W^K_4)^T}{8 }) W^V_4 ]_{N\times 64} \cdot W^{\prime  O}_{64 \times 64}~,
\end{align}
where $W^{Q,K,V}$ is constructed from $M_{N\times 64} \cdot W^{\prime Q,K,V}_{64 \times 16}$, and $W^{\prime Q,K,V,O}$ are trainable parameter matrices.

\begin{figure}[htbp]
\includegraphics[width=0.45\textwidth]{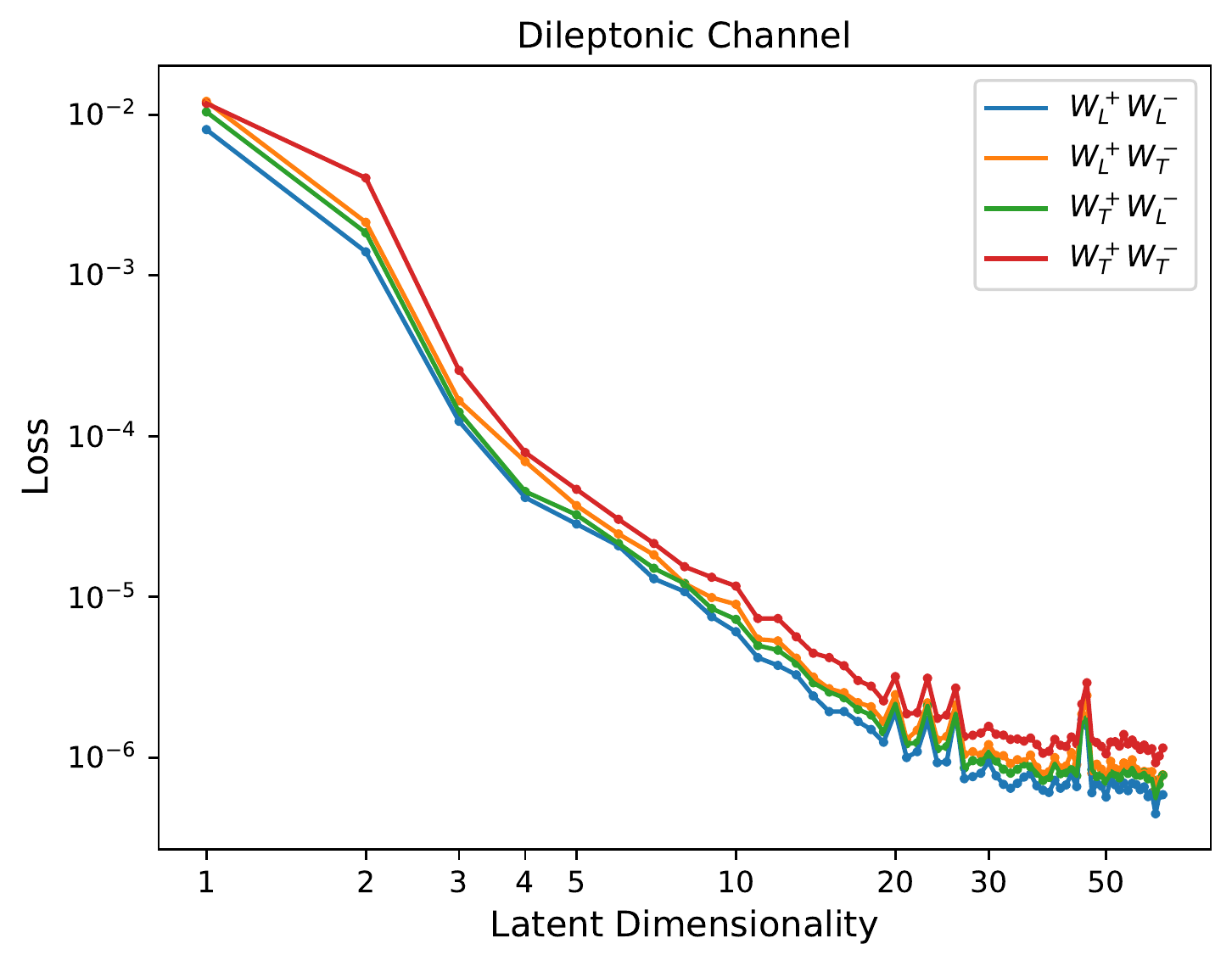}
\includegraphics[width=0.45\textwidth]{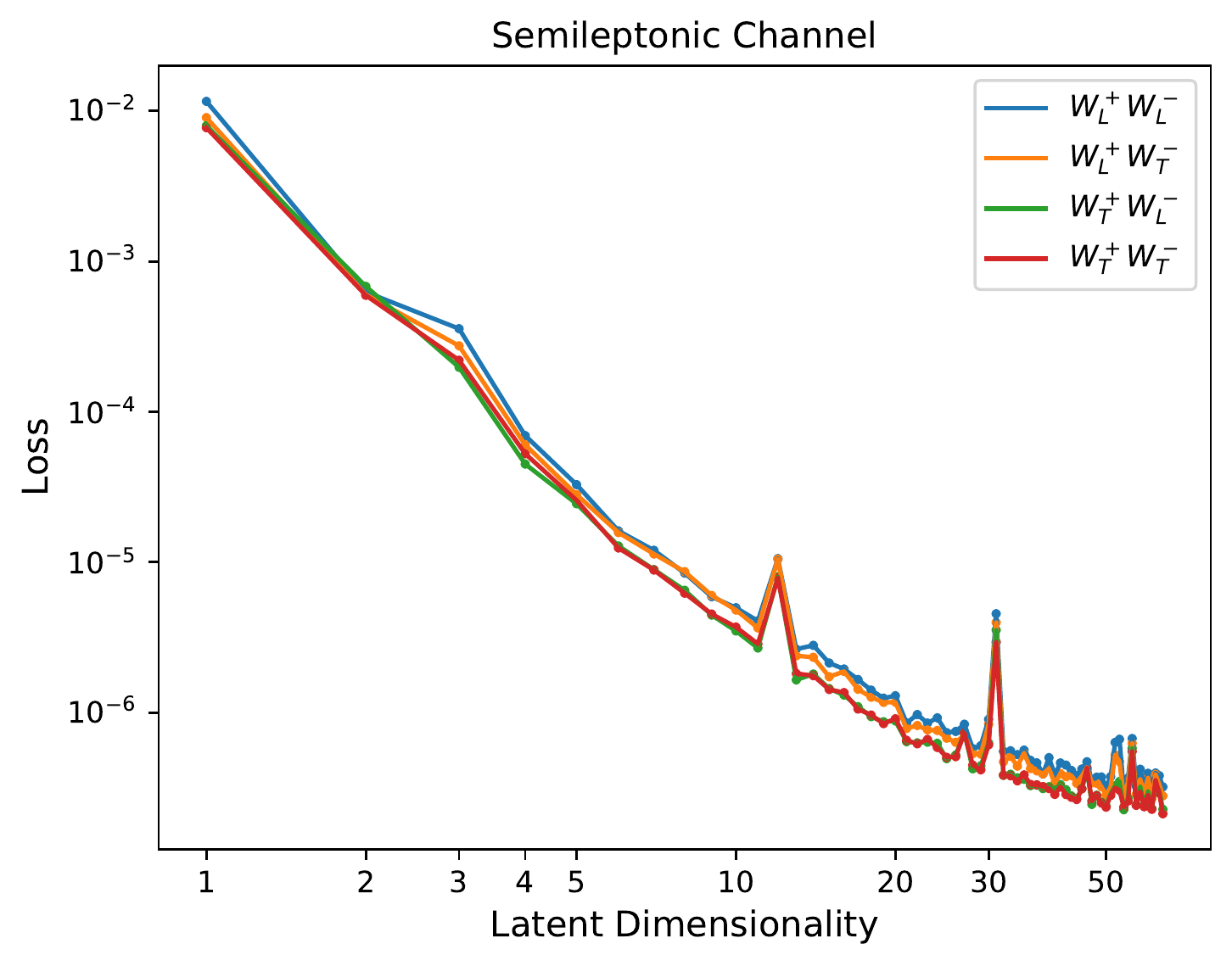}
\caption{Stabilized loss of the Autoencoder for different choices of the dimensionality of feature space in dileptonic channel (left panel) and semi-leptonic channel (right panel).  \label{fig:autoloss}}
\end{figure}

The output of the Transformer is a matrix of size $N\times64$.
The features are obtained by averaging over the particle index (thus it has the shape $1 \times 64$).
Eventually, a Classifier and an Autoencoder is applied for classifying the inputs (to the processes which they belong) and reducing the dimensionality of the feature space. 
The Classifier and Autoencoder are trained simultaneously, using the Adam optimizer with learning rate of $3 \times 10^{-4}$.
Even though higher dimensional feature space provides better discrimination power, the statistical uncertainty in shape analysis is significantly larger due to the limited number of simulated events ($\mathcal{O}(10^{5})$ for each signal processes after preselection). In Fig.~\ref{fig:autoloss}, we show the stabilized loss (typically after $\sim $100 epochs in the training) of the Autoencoder for different choices of the dimensionality of feature space. We can find that for all of the polarization modes in both dileptonic and semi-leptonic channels, the three dimensional latent space can reproduce the 64-dimensional features reasonably well (with loss $\lesssim 10^{-4}$). Meanwhile, the binned log-likelihood analysis can be performed with relatively small statistical uncertainty.

\subsection{Binned log-likelihood analysis in the latent space} \label{sec:likelihood}

The 3-dimensional latent space is divided into $8 \times 8 \times 8$ bins for  dileptonic channel and $10 \times 10 \times 10$ bins for semi-leptonic channel, since the latter has larger production rate.
In principle, one could perform the binned log-likelihood test over all of the bins. However, we find this renders the result sensitive to the tail of the distribution, where the signal and background event numbers are small. Although more dedicated analysis is possible to resolve this issue, we try to only use bins that contain relatively large number of signal events as a simpler alternative.
Among the bins which contain at least 1\% of total signal events, ten with highest signal to background ratios are selected for the log-likelihood test~\footnote{For the EFT case, since the kinematic feature of $W^{\pm}W^{\mp}jj$ production with non-zero $\bar{c}_H$ is similar to that of the SM $W^{\pm}W^{\mp}jj$, the selected bins are identical in most of the cases. As for the 2HDM, around half of the selected bins are different from those of SM $W^{\pm}W^{\mp}jj$. Moreover, the selected bins are different from parameter point to parameter point in the 2HDM. }.
The backgrounds here refers to the summed contributions of $t t_\ell$, $t W_\ell$/$t_\ell W$, $W_\ell Wjj^{\text{QCD}}$, $W_\ell Z jj^{\text{QCD}}$ and $W_\ell Z jj^{\text{EW}}$ processes.
And the signal here refers to the $W_\ell W jj^{\text{EW}}$ and its new physics modifications.
In realistic experiments, the number of signal in each bin can be obtained by subtracting the predicted background event number from the measured number.
This procedure selects $\sim$ 30\% of signal events and $\sim$ 0.5\% of total background events in most of the cases. According to the cross sections in the Tab.~\ref{tab:xsec}, this procedure reduce the cross section of combined backgrounds to the same level as that of the VBS signal.

For a given hypothesis $\mathcal{H}$ (either the SM or new physics BSM), the expected number of events ($t_i$) in the $i$th bin can be obtained from Monte Carlo simulation.
The probability of the $i$th bin having $n_i$ observed events follows the Poissonian probability, $t_i^{n_i} e^{-t_i} / n_i !$.
So we can determine the probability of the full distribution by multiplying the Poissonian probabilities of the selected bins. The binned likelihood for hypothesis $\mathcal{H}_\alpha$ is defined as
\begin{align}
\mathcal{L}(\text{data}| \mathcal{H}_\alpha) = \prod_{i} \frac{t_i^{n_i} e^{-t_i} }{n_i!},
\end{align}
where $i$ runs over 10 selected bins.
Subsequently, we can define the test statistic $\mathcal{Q}$ as the log likelihood ratio between a given hypothesis $\mathcal{H}_\alpha$ (new physics with fixed parameters) and the null hypothesis $\mathcal{H}_0$ (the SM).
\begin{align}
\mathcal{Q} = -2 \log \left( \frac{\mathcal{L}(\text{data}| \mathcal{H}_\alpha)}{\mathcal{L}(\text{data}| \mathcal{H}_0 )}  \right).
\end{align}
We use the expected numbers of events from two hypotheses ($\mathcal{H}_\alpha$ and $\mathcal{H}_0$) to generate two sets of pseudo-data.
In each bin, the pseudo-data is obtained by generating a random number from Poissonian (statistical uncertainty) plus Gaussian distribution (systematical uncertainty) with mean value of $t_i$. We repeat this procedure $10^6$ times for $\mathcal{H}_\alpha$ and $\mathcal{H}_0$, respectively. This gives two distributions of the test statistic $\mathcal{Q}$.
Finally, the $p$-value of the test hypothesis ($\mathcal{H}_{\alpha}$) can be calculated by assuming that the actual observation is at the center of $\mathcal{Q}$ distribution under null hypothesis.

\section{Learning the features of vector boson polarization} \label{sec:pol}
Among polarization modes of the VBS processes, the longitudinally polarized component is most closely related to the unitarity issue, i.e. the property of the Higgs boson and possible new physics.
There have been extensive studies on separating the polarization of the gauge boson in the VBS process, exploiting various kinematic variables.
The lepton angular distribution in the gauge boson rest frame is known to be sensitive to the vector boson polarization,
\begin{align}
\frac{1}{\sigma} \frac{d \sigma}{d \cos \theta} &= \frac{3}{8} f_L (1+\cos \theta)^2 + \frac{3}{8} f_R (1-\cos \theta)^2
  + \frac{3}{4} f_{0} \sin^2 \theta,  ~\label{eq:pol}
\end{align}
where the $f_{L,R,0}$ is the fraction of the corresponding helicity and the $\theta$ is the angle between the vector boson flight direction in a certain frame and the lepton flight direction in the vector boson rest frame.
Even though the shape of the angular distribution is a good discriminating variable, it can not be reconstructed precisely for the most of the time.
In the dileptonic channel of $W^\pm W^\mp jj$, there are two missing neutrinos in the final state. One can not reconstruct the rest frame for individual $W$ boson.
As for the semi-leptonic channel, even though the neutrino momentum can be solved up to a twofold ambiguity (thus the full momenta of all particles can be calculated), there are usually large uncertainties in measuring the jets momenta and identifying the forward-backward jets and jets from $W$ boson decay.
Moreover, the shape of the $\theta$ distribution can be distorted by kinematic cuts that need to be used to separate VBS from its backgrounds~\cite{Stirling:2012zt}.

In this section, we demonstrate that our network is capable of discriminating different polarization modes of the electroweak $W^\pm W^\mp jj$ production with the low-level inputs.

\subsection{The dileptonic channel}

\begin{figure}[htbp]
\includegraphics[width=0.24\textwidth]{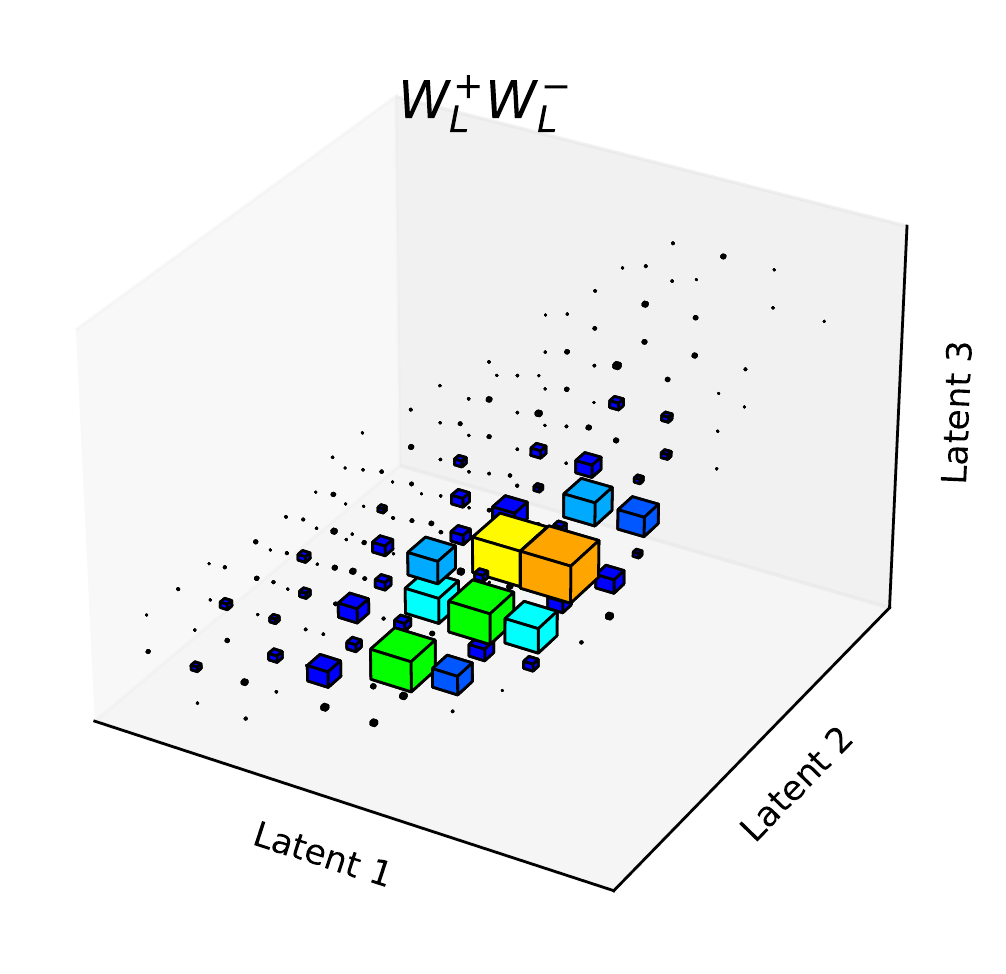}
\includegraphics[width=0.24\textwidth]{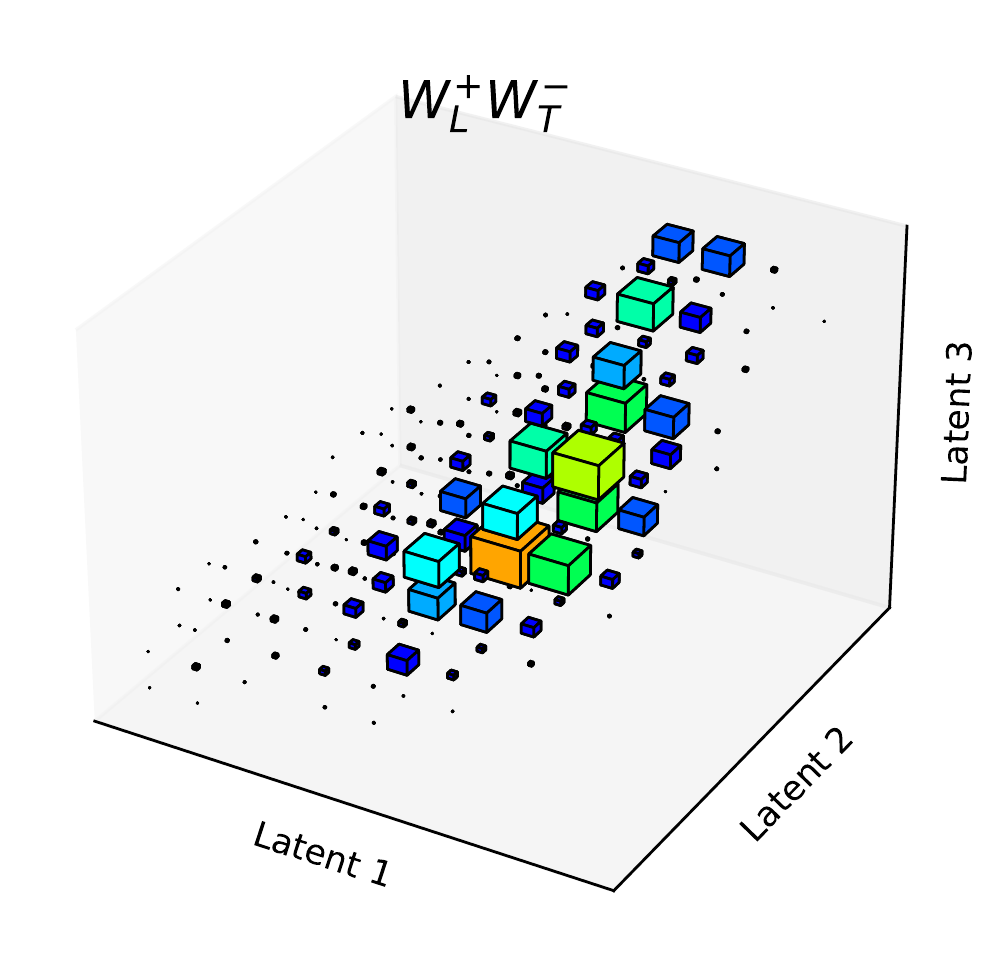}
\includegraphics[width=0.24\textwidth]{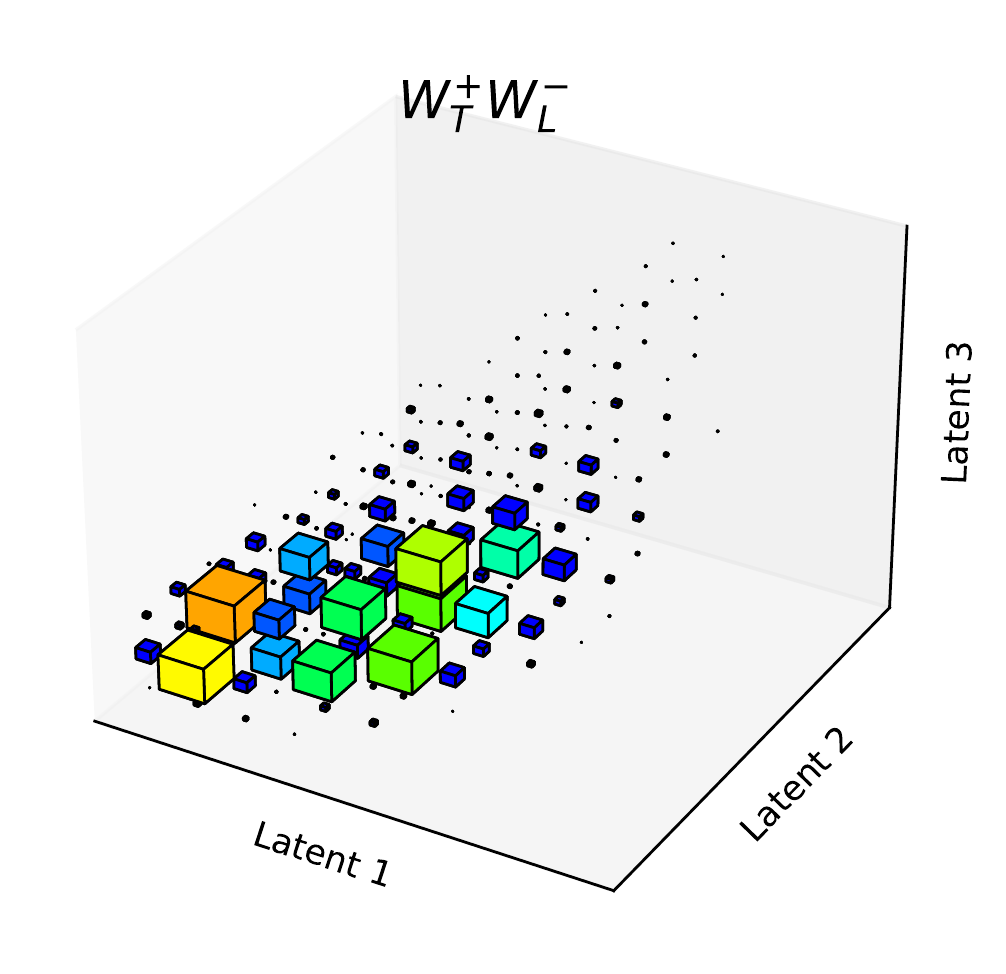}
\includegraphics[width=0.24\textwidth]{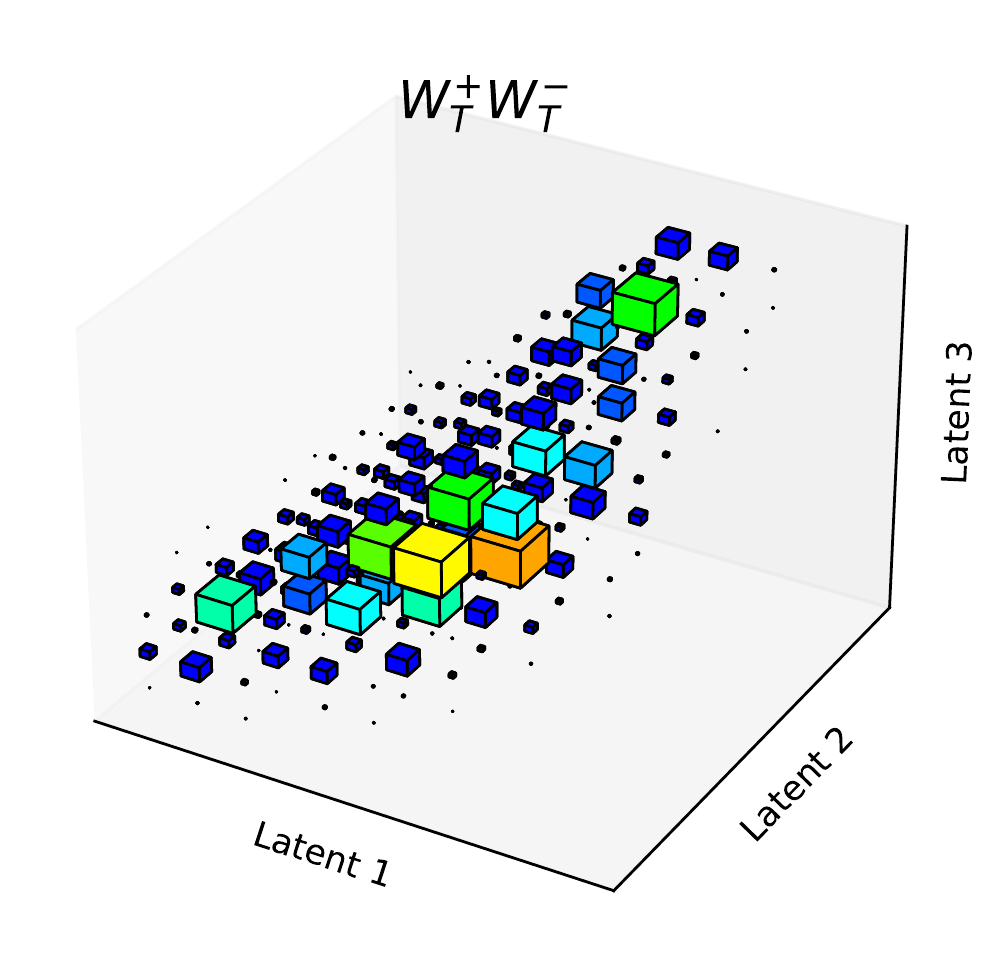}
\caption{The normalized distributions of the latent features for different polarization modes in the dileptonic channel.  \label{fig:3dll}}
\end{figure}

We train the network with labeled events of electroweak $W^+_L W^-_L jj$, $W^+_L W^-_T jj$, $W^+_T W^-_L jj$, $W^+_T W^-_T jj$ productions, respectively. Here $W_L$ ($W_T$) represents longitudinally (transversely) polarized $W$ boson.
The normalized~\footnote{Integrating the distribution over all bins gives one. } distributions of those polarization modes in the three dimensional latent space are shown in Fig,~\ref{fig:3dll}. Larger size of cube indicates more events in that bin. We can find remarkable differences in the distributions of different polarizations.

To assess the discriminating power of our network, we perform a comparative study on methods with different input variables.
Besides the three latent features, two classes of variables are defined~\footnote{We have tried many other variables, only those showing significant discriminating power are kept.}:
\begin{itemize}
\item {\textbf{Detector level variables:}} Variable in this class can be reconstructed experimentally, include the transverse momenta of two leptons $p_T (\ell_{1,2})$ and the forward-backward jets $p_T(j_{1,2})$; the azimuthal angle difference between the forward and backward jets $\Delta \phi(j,j)$.
\item {\textbf{Truth level variables:}} Variable in this class can only be obtained from Monte Carlo simulation, include the transverse momenta of two $W$ bosons $p_T(W^\pm)$; the lepton angle in the $W$ boson rest frame $\cos(\theta_{l^\pm})$. The later is calculated by $\cos \theta = \frac{\vec{p}_W \cdot \vec{\tilde{p}}_\ell}{|\vec{p}_W|  |\vec{\tilde{p}}_\ell |}$, where $\vec{p}_W$ is the $W$ boson momentum in the initial parton center of mass frame and $\vec{\tilde{p}}_\ell$ is the lepton momentum in the $W$ boson rest frame.
\end{itemize}

\begin{figure}[htbp]
\includegraphics[height=6cm,width=8cm]{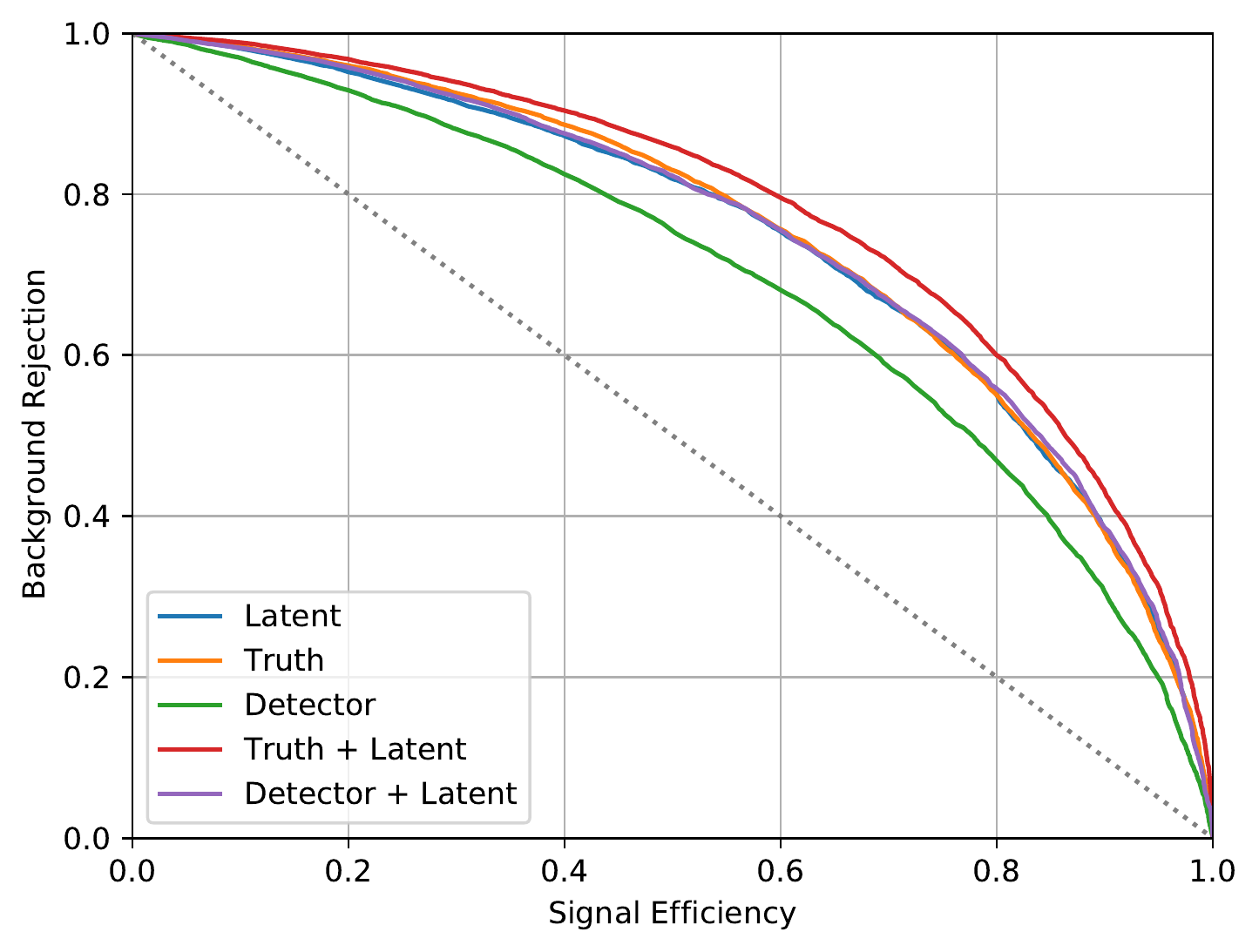}
\includegraphics[height=6cm,width=8cm]{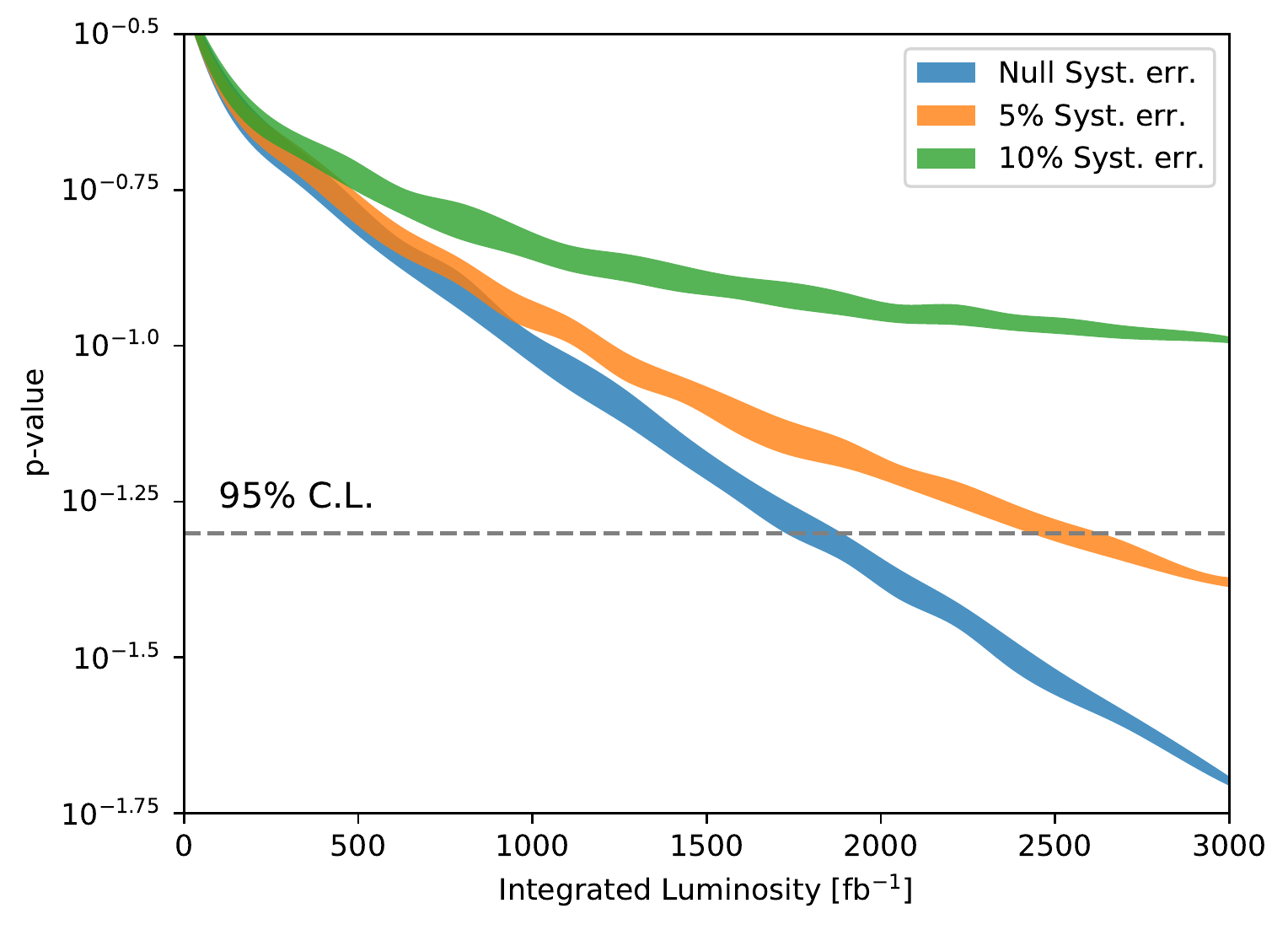}
\caption{Left: comparison of the discriminating power of methods with different input variables in the dileptonic channel. Right: the sensitivity to 1\% change in the rate of the $W^+_L W^-_L jj$ mode. The width of a band indicates the statistical uncertainty and the color represents different systematic uncertainties. \label{fig:rocll}}
\end{figure}

The Gradient Boosting Decision Tree (GBDT) method is adopted to calculate the receiver operating characteristic (ROC) curves with inputs of the variables in a class either with or without including the latent variables. The ROC curves are showing in the left panel of Fig.~\ref{fig:rocll}, where we have considered the events of the $W^+_L W^-_L jj$ as the signal and events of other polarization modes as background.
We can find that the method using latent features alone have already outperform the GBDT with all detector level variables. And the GBDT which combines the latent variables with the detector level variables does not have better discriminating power than the method with solely latent variables. It indicates that the information of those detector level variables should have been included in the latent variables.
The GBDT with truth level variables have slightly improved discriminating power than the method with latent variables. It is also interesting to observe that the discriminating power can be improved further by combining the truth level variables and latent variables.

When the new physics modifies the Higgs to gauge boson interaction, the incomplete cancellation in the VBS amplitude leads to an increased fraction of longitudinal polarized gauge boson final state. The current precision measurements of the SM allows the increasement of $W^+_L W^-_L jj$ fraction by a percent level, e.g. from 6\% to 7\% in the following case.
To study the sensitive of latent variables to this amount of change, we perform the binned log-likelihood analysis, taking the SM cross section (after applying the cut of $m_{jj}>500$ GeV at parton level) for each polarized component. These are $\sigma(W^+_L W^-_L)=25.5$ fb, $\sigma(W^+_L W^-_T)=73.2$ fb, $\sigma(W^+_T W^-_L)=76.9$ fb and $\sigma(W^+_T W^-_T)=243.8$ fb respectively. The test hypothesis take $\sigma(W^+_L W^-_L)=29.7$ fb while keep other cross sections the same.
The p-values for the hypothesis test with varying the integrated luminosity are shown in Fig.~\ref{fig:rocll}, where we have considered the cases with three different systematical uncertainties. We can conclude that future LHC is capable of detecting such change, if the systematic uncertainty is below $\sim 5\%$.
Note that the background processes are not considered at this stage. Moreover, the new physics may not be simply considered as the summation of the SM components. More complete and realistic analysis will be given in the next two sections.

\subsection{The semi-leptonic channel}

\begin{figure}[htbp]
\includegraphics[width=0.24\textwidth]{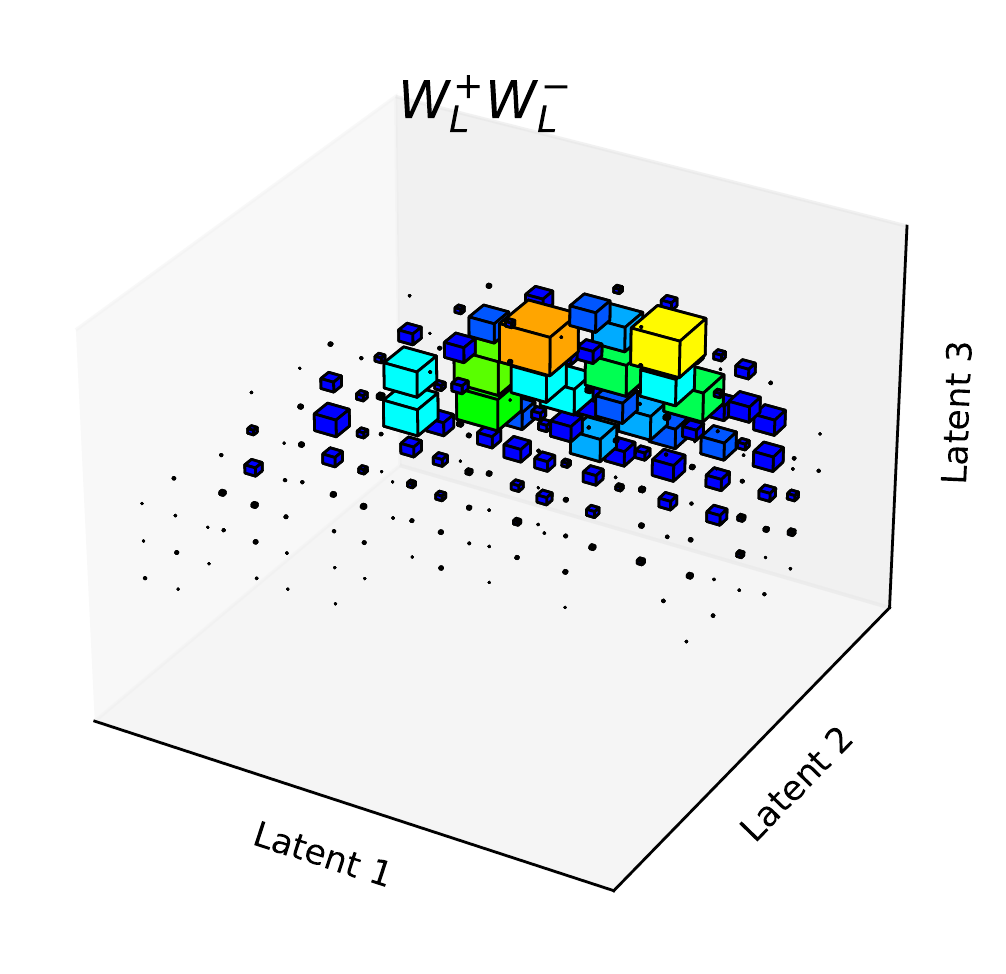}
\includegraphics[width=0.24\textwidth]{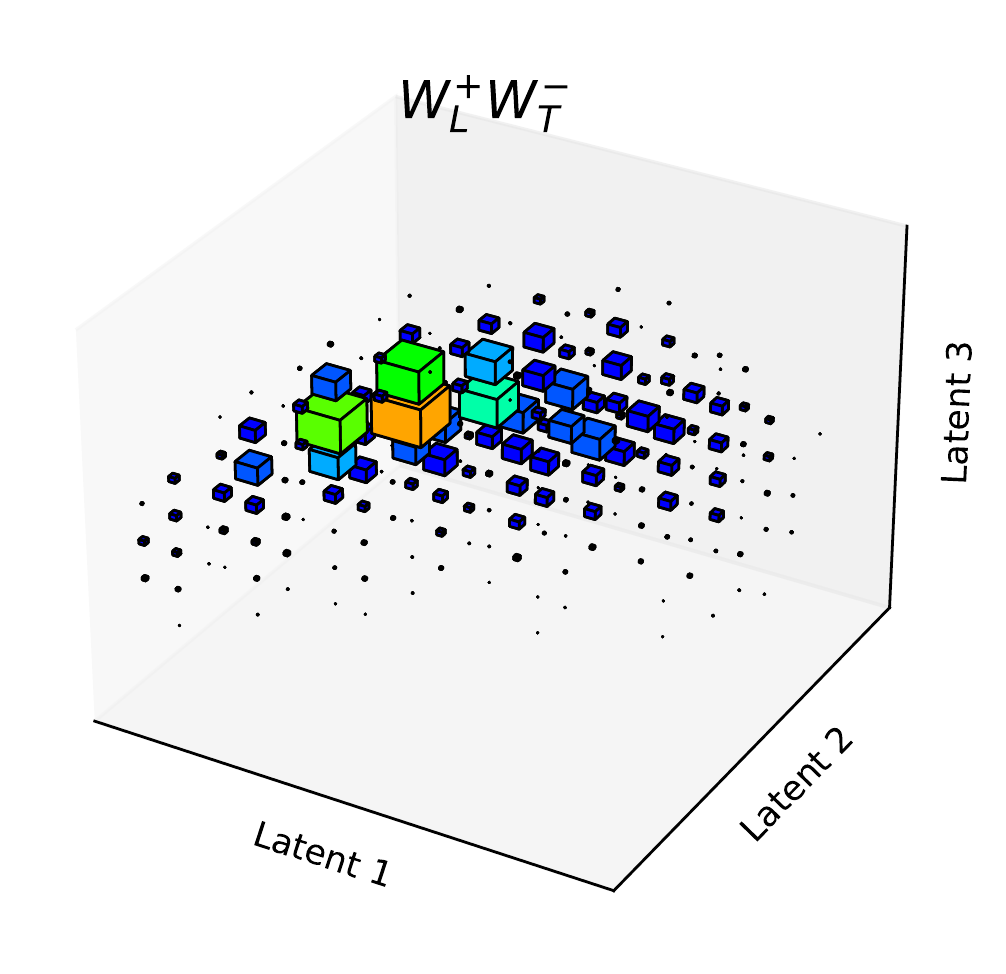}
\includegraphics[width=0.24\textwidth]{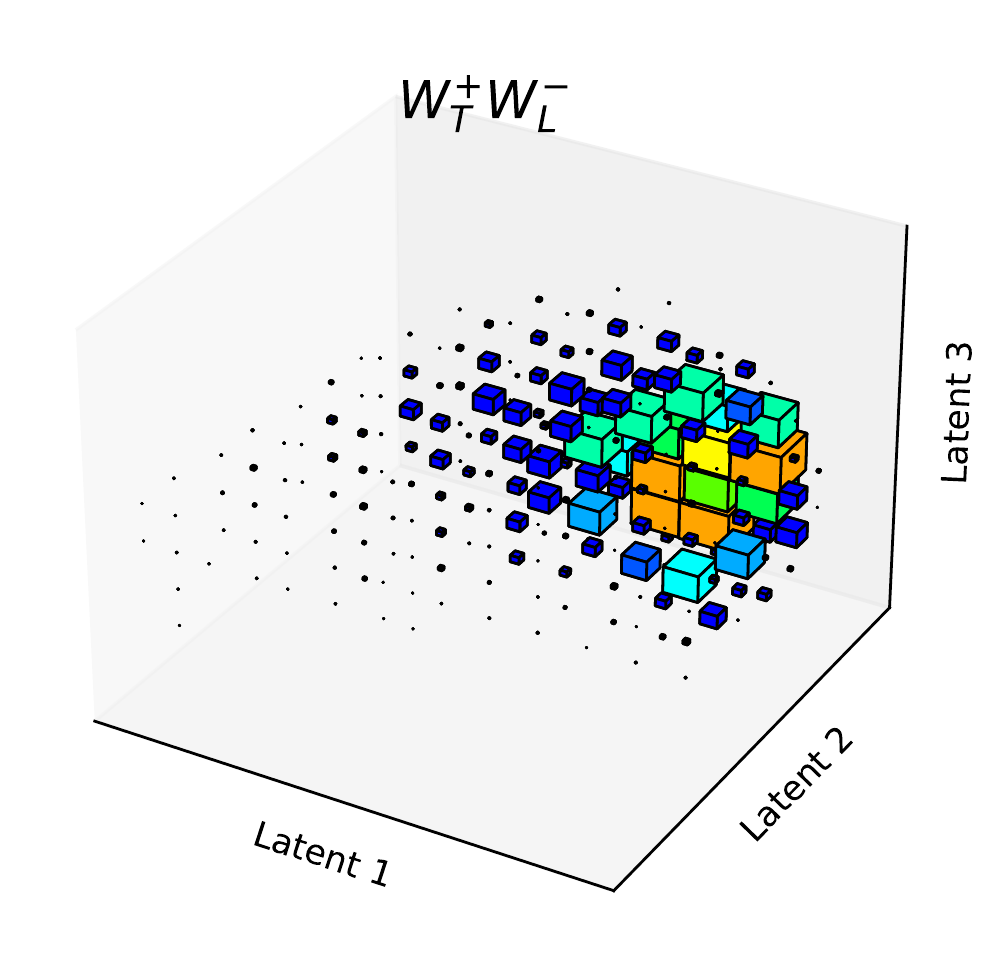}
\includegraphics[width=0.24\textwidth]{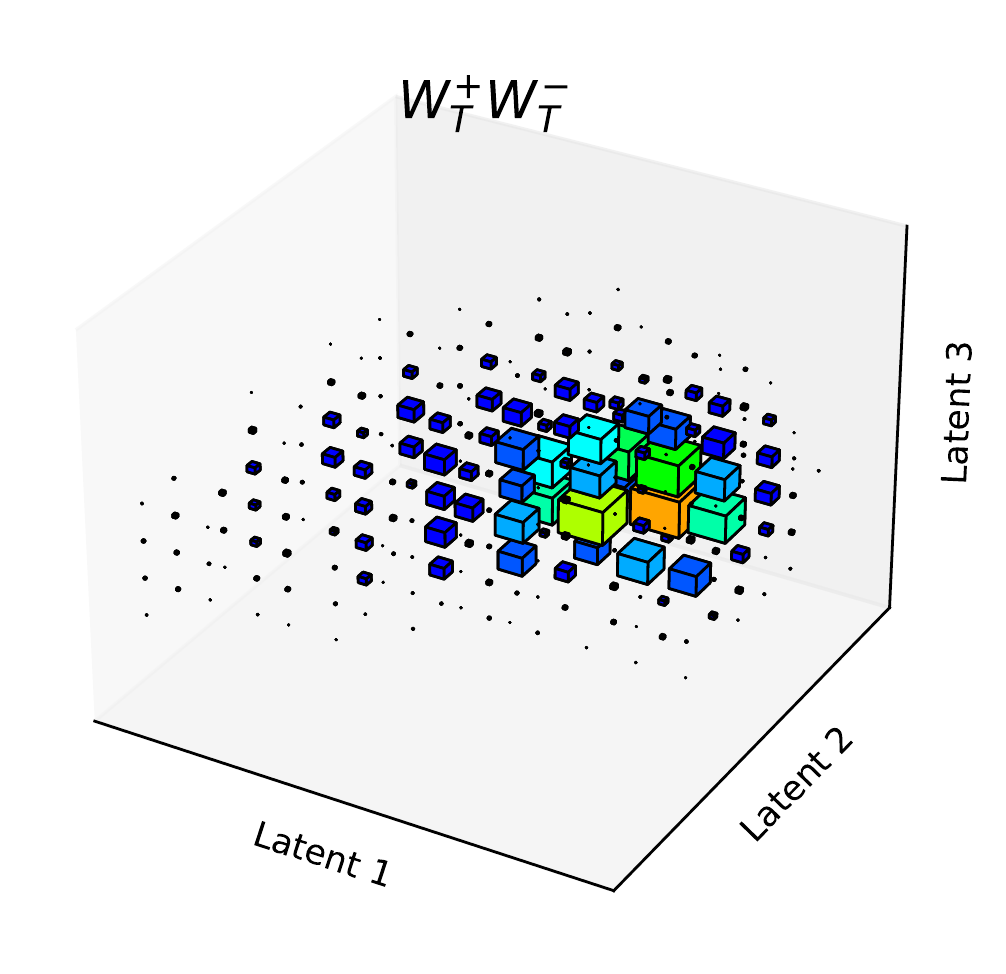}
\caption{The normalized distributions of the latent features for different polarization modes in the semi-leptonic channel.  \label{fig:3dlj}}
\end{figure}

Comparing to the dileptonic channel, the semi-leptonic channel has much larger production cross section and only includes a single neutrino in the final state. Better discrimination power can be obtained in this channel.
Similarly, the network for the semi-leptonic channel is trained with labeled events of EW production of $W^\pm W^\mp jj$ with different polarizations. The normalized distribution for each polarization mode in the latent space is shown in Fig.~\ref{fig:3dlj}.

Two classes of variables that are used in GBDT method to calculate the ROC curves are listed as follows.
\begin{itemize}
\item {\textbf{Detector level variables:}} transverse momentum $p_T(\ell)$ and pseudorapidity $\eta(\ell)$ of the lepton,  azimuthal angle difference between forward backward jets $\Delta \phi (j,j)$ and the transverse momentum of $W$ boson pair $p_T(W,W)$ which can be calculated by vector sum of the transverse momenta of its decay products (including the missing transverse momentum).
\item {\textbf{Truth level variables:}} transverse momenta of two $W$ bosons $p_T(W^\pm)$, the lepton angle in the W boson rest frame $\cos(\ell)$ and the invariant mass of the forward backward jets $m_{jj}$.
\end{itemize}

\begin{figure}[htbp]
\includegraphics[height=6cm,width=8cm]{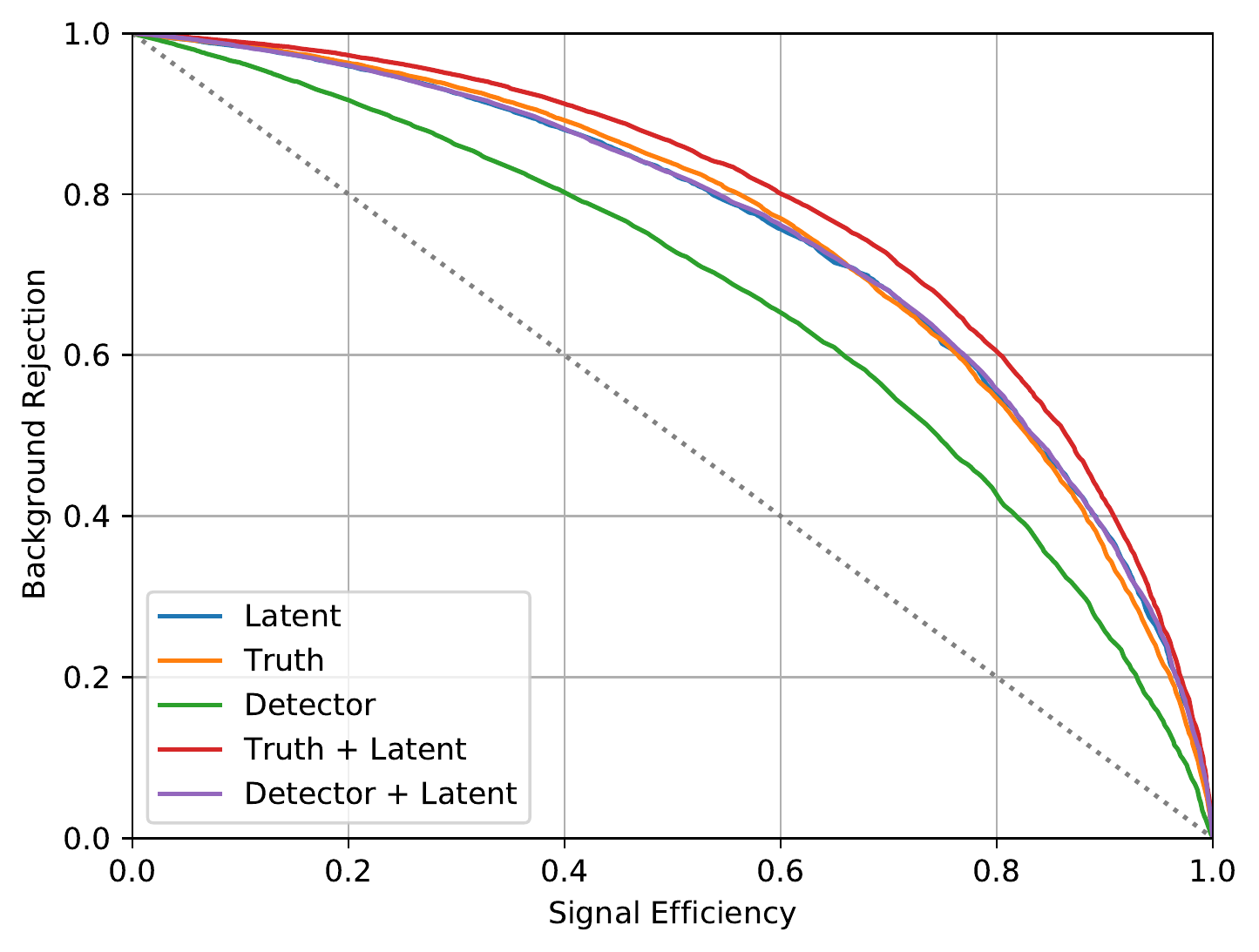}
\includegraphics[height=6cm,width=8cm]{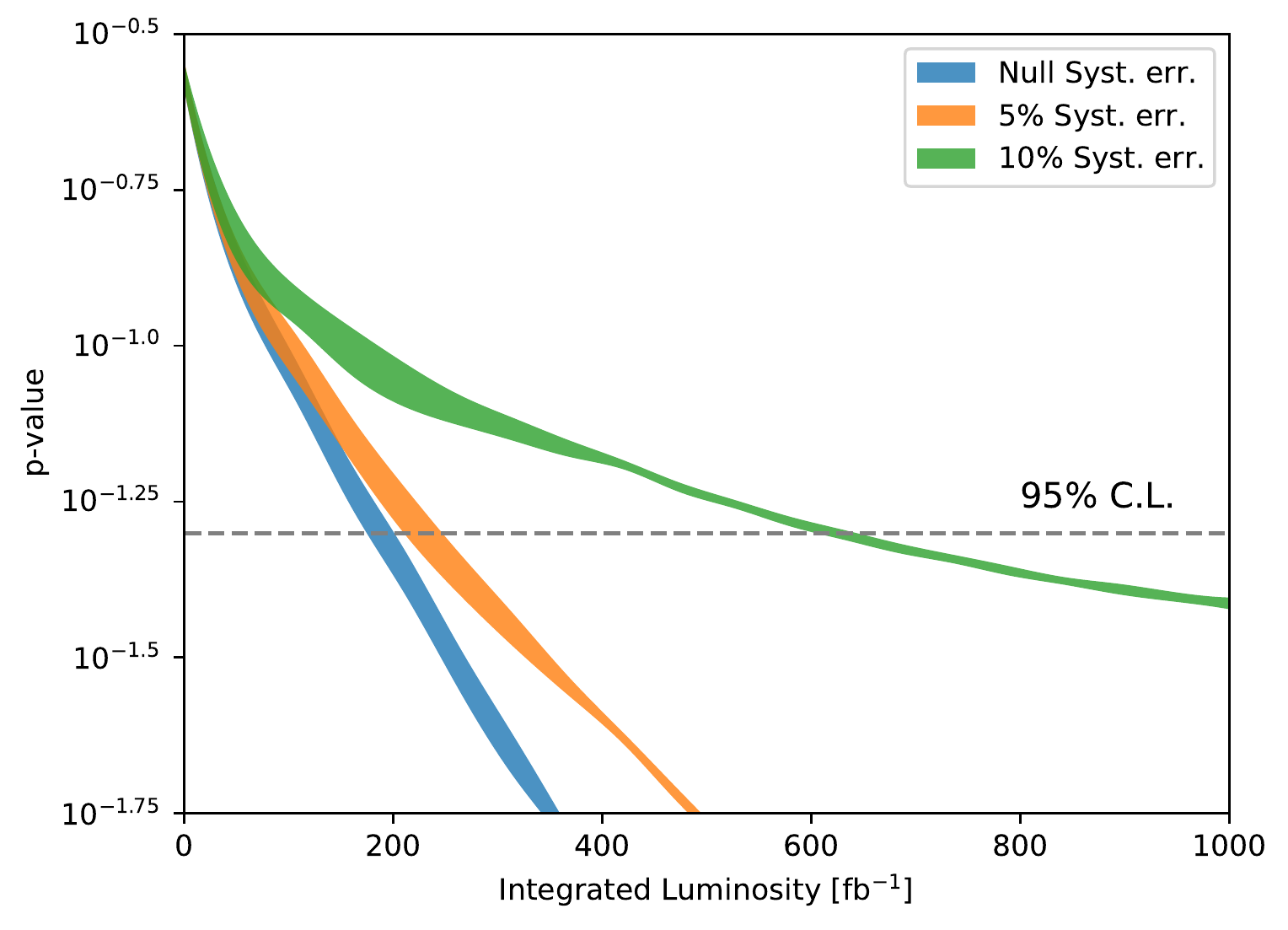}
\caption{Same as Fig.~\ref{fig:rocll}, but for the semi-leptonic channel. Note the variables used for plotting the ROC is different from that in the dileptonic channel. \label{fig:roclj}}
\end{figure}

The ROC curves for methods with different inputs are presented in the left panel of Fig.~\ref{fig:roclj}.
Even though the semi-leptonic channel only contain one neutrino in the final state, the large uncertainty in jet measurement and confusion of forward-backward jets with jets from $W$ boson decay render the similar polarization discriminating power of this channel with that of the dileptonic channel.
However, due to the sizable production rate of this channel, dataset with integrated luminosity of $\lesssim 600$ fb$^{-1}$ can be used to probe the 1\% change in the $W^+_L W^-_L jj$ fraction.

It should be noted that this result is only provided as a rough estimation. In a concrete model, the differential cross section of the EW $W^+ W^- j j$ channel is not simply given by the combination of the SM polarization components. Variables other than those listed above can be helpful in discriminating different polarizations. Meanwhile, the contribution from SM background processes should be taken into account.
In the following two sections, we will consider the effective field theory and two-Higgs-Doublet-Model (2HDM) as case study.

\section{Application to the effective field theory}\label{sec:eft}
In absence of direct observations of new states, a practical way for investigating the new physics lies in a description based on the EFT, which is valid up to the scale of new physics.
The EFT contains a complete set of independent gauge-invariant operators made up by the SM fields.
There have been numerous studies on constraining the coefficients of these operators with precision measurements at experiments~\cite{DiVita:2017eyz,Ellis:2018gqa,Grojean:2018dqj,Biekotter:2018rhp,Almeida:2018cld}.
Most of the operators are tightly constrained by the elctroweak precision tests (EWPT) of the SM.
We will consider the following operator~\cite{Giudice:2007fh,Contino:2013kra}
\begin{align}
\mathcal{O}_H = \frac{\bar{c}_H}{2 v^2} \partial^\mu [\Phi^\dagger \Phi]  \partial_\mu [\Phi^\dagger \Phi] \Rightarrow  \frac{\bar{c}_H}{2} \partial^\mu h \partial_\mu h
\end{align}
since it is less constrained by the EWPT. The $\Phi$ field is Higgs doublet and $h$ denotes the Higgs boson field with the vacuum expectation value $v=246.2$ GeV.
The $\mathcal{O}_H$ operator contributes to the Higgs boson kinetic term, and an appropriate field redefinition is required to bring back the kinetic term to its canonical form
\begin{align}
h \to h[1 - \frac{1}{2} c_H].
\end{align}
It leads to the following changes to the Higgs couplings
\begin{align}
\mathcal{L}_{H}  \supset & \frac{g m_W}{c^2_W} [1- \frac{1}{2} \bar{c}_H]  Z_\mu Z^\mu h  + g m_W [1-\frac{1}{2}\bar{c}_H ] W^\dagger_\mu W^\mu h  \nonumber \\
 & + [ \frac{y_f}{\sqrt{2}} [1-\frac{1}{2}\bar{c}_H] \bar{f} P_R f h +h.c.]~. \label{eq:hcouplings}
\end{align}
The up-dated global fit to the EFT coefficients constrains $\bar{c}_H \lesssim 0.4$ (marginalizing over all other operators)~\cite{Dawson:2020oco}. Future lepton colliders, such as the ILC, will constrain the $\bar{c}_H$ to the 1\% level~\cite{Jung:2020uzh}.

We study its effects on the EW $W^+ W^- jj$ production at the LHC.
As the polarization vector $\epsilon_L^\mu \sim \frac{p^\mu}{m_V} + \mathcal{O}(\frac{m_V}{E})$ grows with momentum $p$, the longitudinally polarized gauge boson scattering ($W_L W_L \to W_L W_L$) is dominant at high energy.
In the high energy limit, the amplitude for the longitudinal $W$ boson scattering without Higgs contribution is
\begin{align}
\mathcal{M}^{\rm gauge} = -\frac{g^2_w}{4 m^2_W} u + \mathcal{O}(s^0)~,
\end{align}
which cancels with the amplitude from Higgs exchange
\begin{align}
\mathcal{M}^{\rm Higgs} = - \frac{g^2_w}{4 m^2_W} [\frac{(s-m^2_W)^2}{s-m^2_H} + \frac{(t-m^2_W)^2}{t-m^2_H}] \stackrel{s,t,u \gg m_W, m_H}{\sim} \frac{g^2_w}{4 m^2_W} u~,
\end{align}
leaving terms not rising with energy.  Here, $s,t,u$ are  Mandelstam variables.
However, the cancellation only holds if the Higgs boson couplings to gauge bosons are exactly SM-like. The $\mathcal{O}_H$ operator modifies the Higgs boson couplings as shown in Eq.~\ref{eq:hcouplings}, leading to an incomplete cancelation up to the scale where new physics states come in. As a result, the fraction of the $W^+_L W^-_L jj$ is increased  and the kinematic properties of final states are changed.

\begin{table}[htb]
\begin{center}
\begin{tabular}{|c||c|c|c|c|c|c|} \hline
$\bar{c}_H$ &  -1.0 & -0.5 & 0 & 0.5 & 1.0 \\ \hline \hline
$\sigma^0_{m_{jj}>500}$ [fb] & 440.6 & 421.8 & 419.7 & 426.7 & 436.2  \\ \hline
$\sigma_{ll}$ [fb] & 4.82  & 4.44 & 4.36 & 4.48 & 4.62  \\ \hline
$\sigma_{lj}$ [fb]  & 40.2 & 37.7 & 37.3 & 37.9 & 39.3 \\ \hline \hline
$\sigma^{LL}_{m_{jj}>500}$ [fb] & 46.29 & 29.68 & 25.84 & 28.79 & 34.01  \\ \hline
$\sigma^{LL}_{ll}$ [fb] & 0.754 &  0.397 &  0.314 &  0.356 &  0.462 \\ \hline
$\sigma^{LL}_{lj}$  [fb] & 5.28 & 3.04 &  2.40 &  2.79 &  3.50  \\ \hline
\end{tabular}
\caption{\label{tab:ch_xsec}   $\sigma^0_{m_{jj}>500}$ and $\sigma^{LL}_{m_{jj}>500}$ are the production cross sections (requiring the invariant mass of forward backward jets to be greater than 500 GeV at parton level) for the total and longitudinal polarized EW $W^+ W^- jj$ productions. $\sigma^{(LL)}_{ll/lj}$ correspond to the cross sections of the dileptonic channel ($ll$) and the semi-leptonic channel ($lj$) after preselection cuts.   }
\end{center}
\end{table}

We adopt the UFO model as implemented in Ref.~\cite{Alloul:2013naa} to generate the EW $W^+ W^- j j$ events in the EFT. All of the coefficients except the $\bar{c}_H$ are set to zero.
Both the dileptonic channel and the semi-leptonic channel are considered.
Only those events that pass through the preselection cuts as listed in Sec.~\ref{sec:events} will be fed into the network for further analyses.
The production cross section of the EW $W^+ W^- j j$ process (with different choices of $\bar{c}_H$) before and after the preselections are given in Tab.~\ref{tab:ch_xsec}.
The $\bar{c}_H=0$ case corresponds to the SM. We can find the fraction of the longitudinal $W$ production increases with $|\bar{c}_H|$ as the cancellation become less exact. And our preselection cuts can raise the fraction of the longitudinal $W^+_L W^-_L j j$, especially for the dileptonic channel.
After the preselections, the production rate of the semi-leptonic channel is an order of magnitude large than that of the dileptonic channel.

\begin{figure}[htb]
\includegraphics[width=0.24\textwidth]{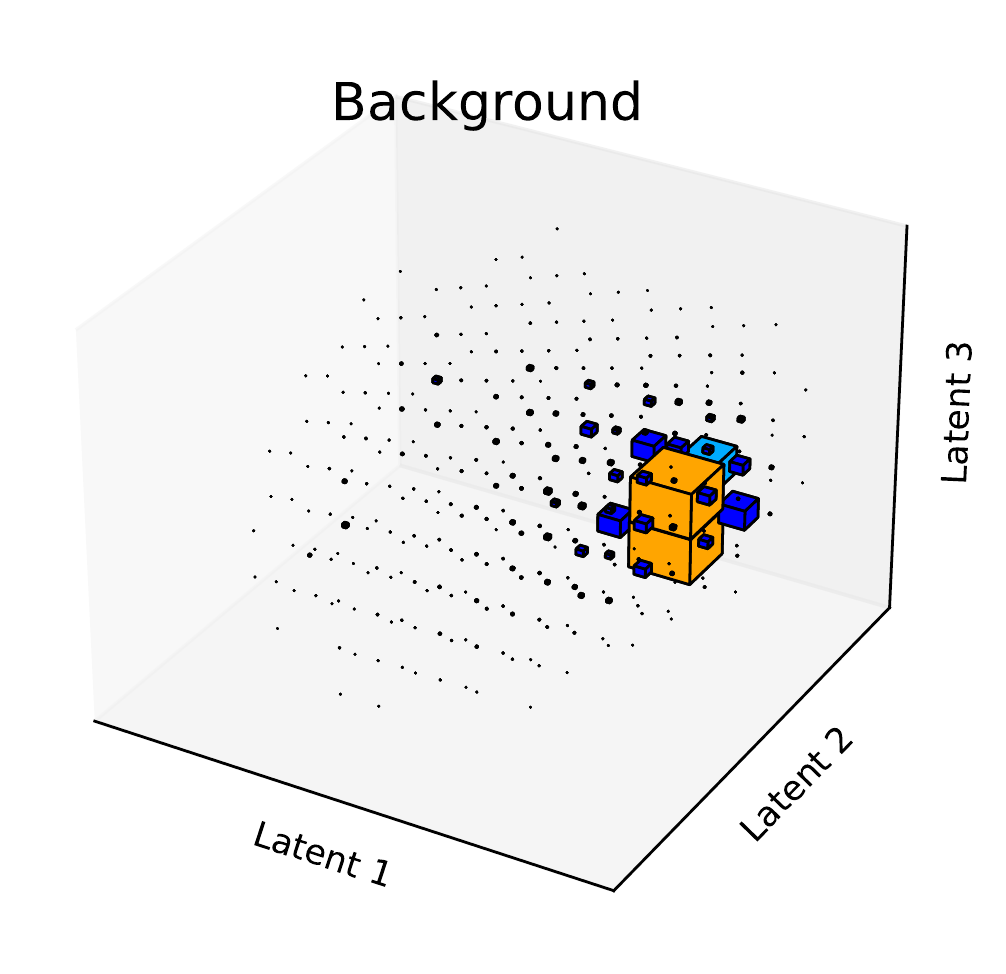}
\includegraphics[width=0.24\textwidth]{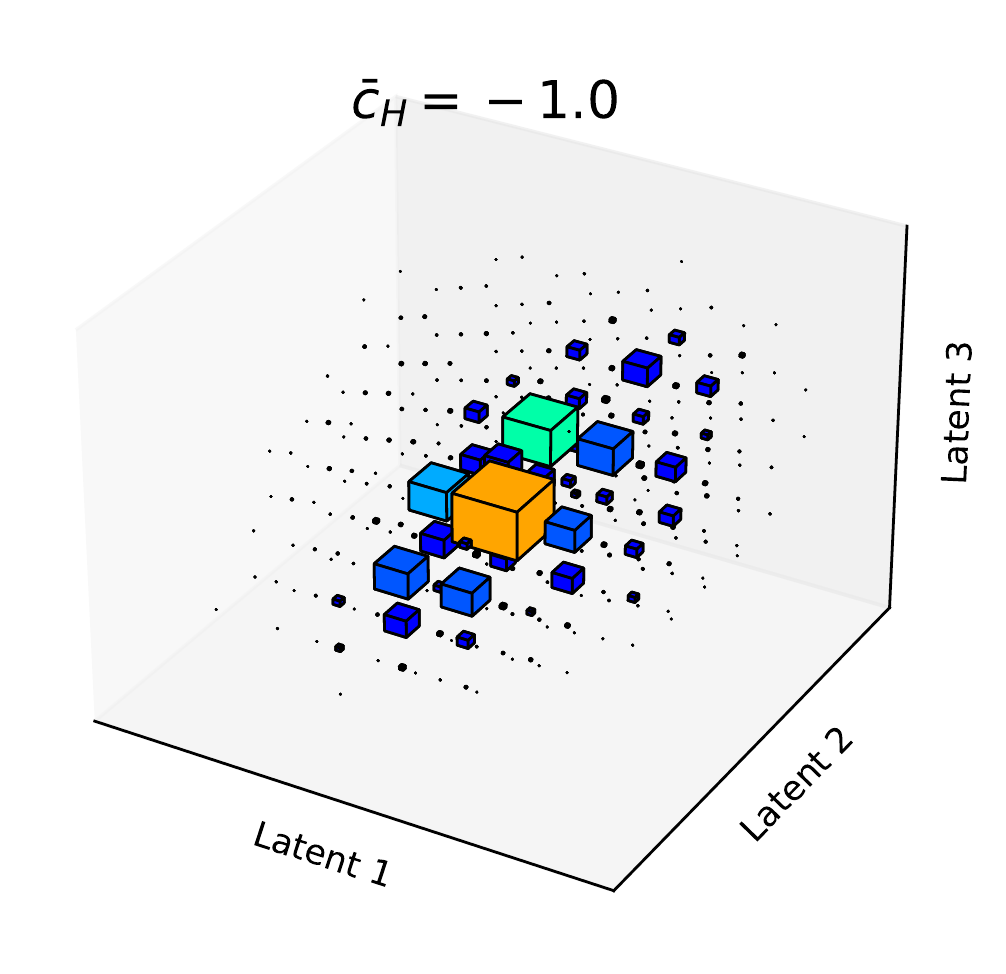}
\includegraphics[width=0.24\textwidth]{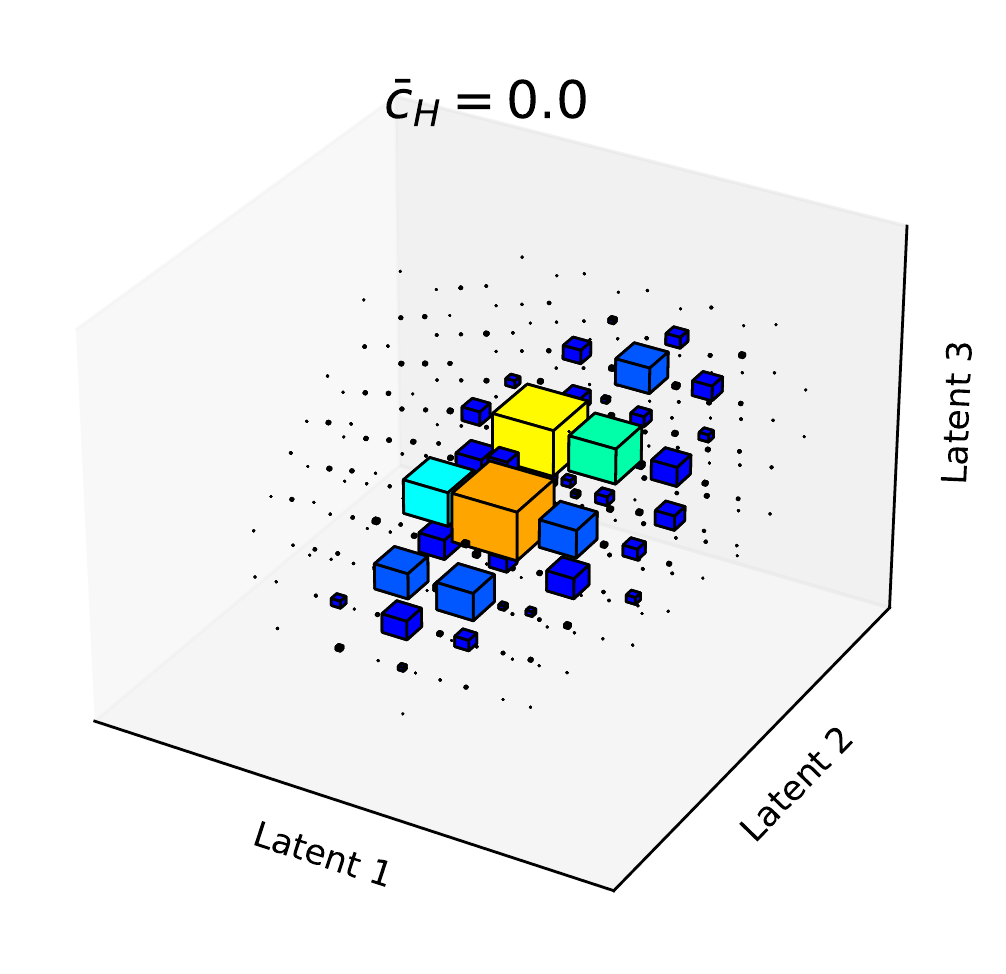}
\includegraphics[width=0.24\textwidth]{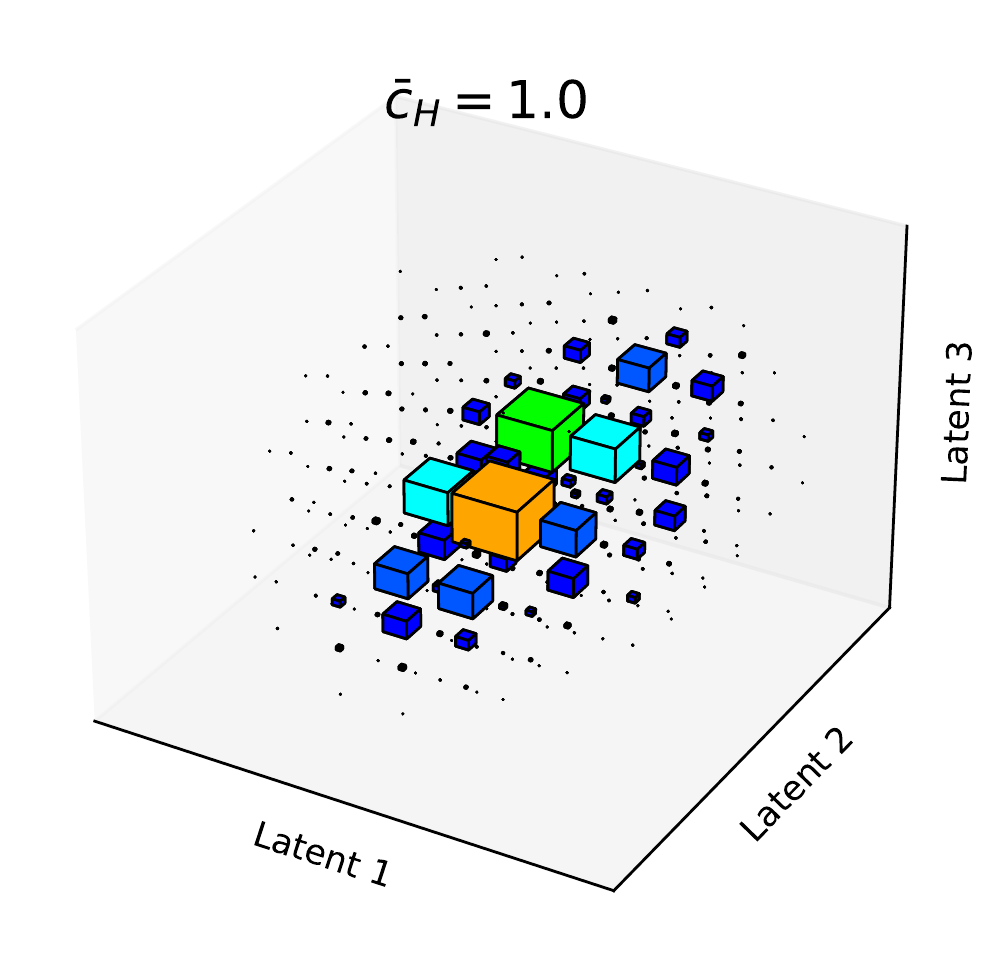} \\
\includegraphics[width=0.24\textwidth]{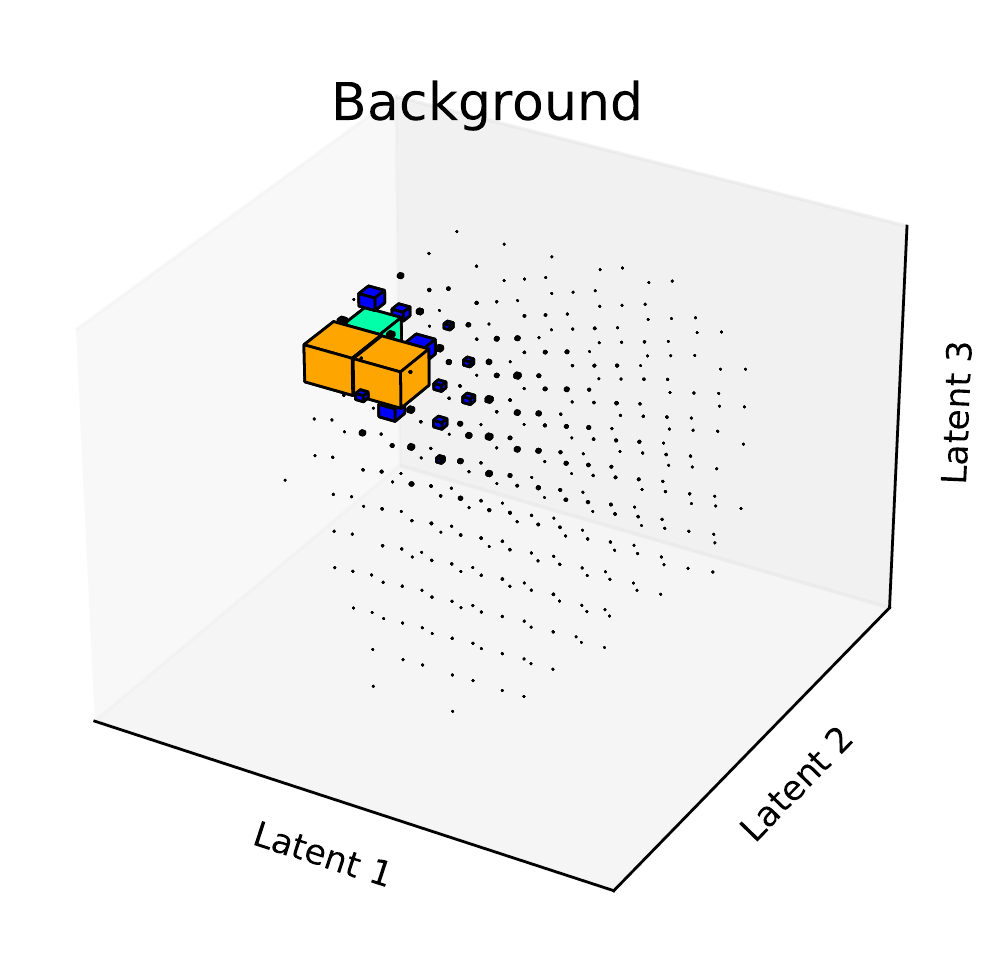}
\includegraphics[width=0.24\textwidth]{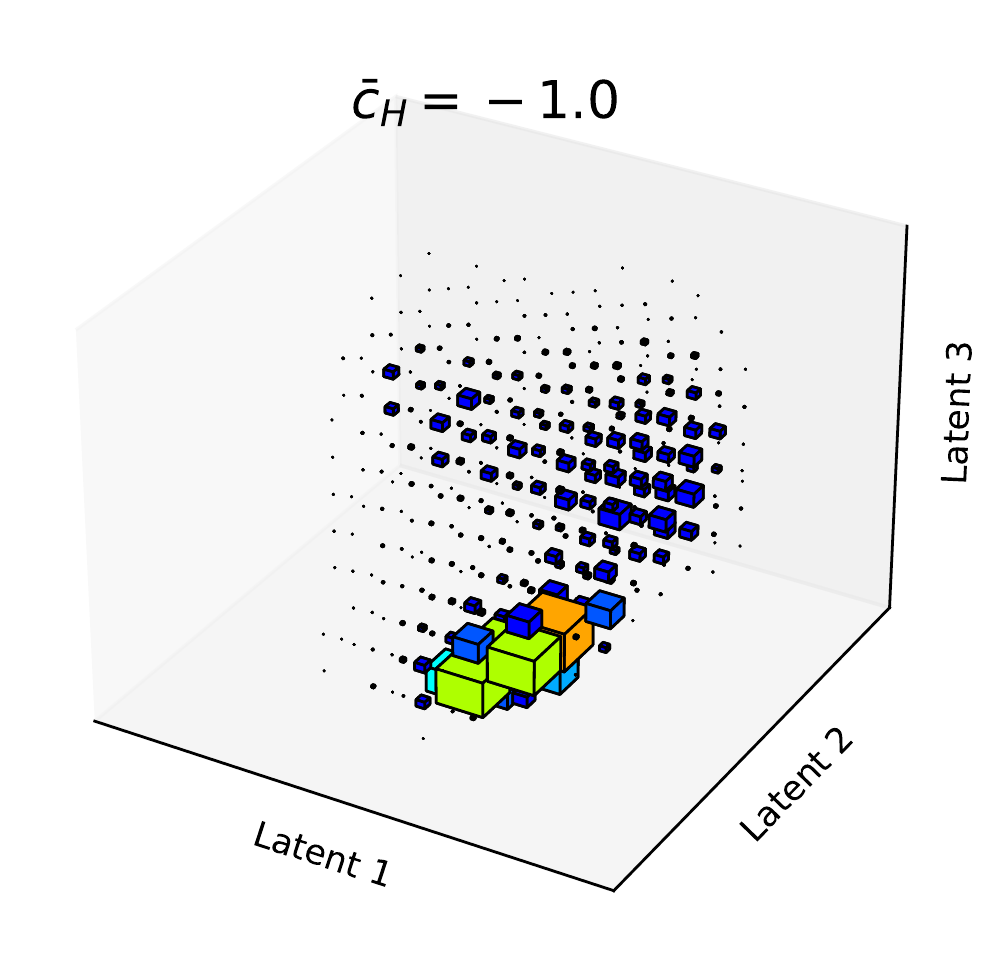}
\includegraphics[width=0.24\textwidth]{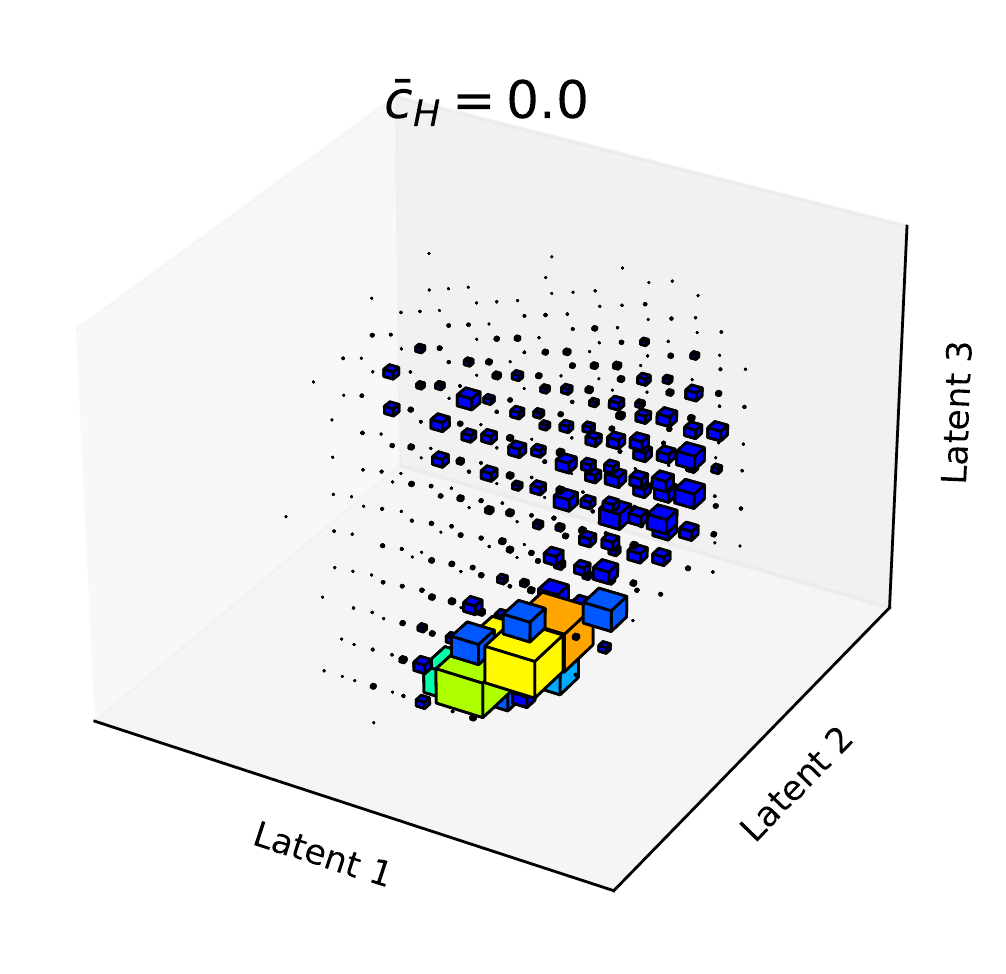}
\includegraphics[width=0.24\textwidth]{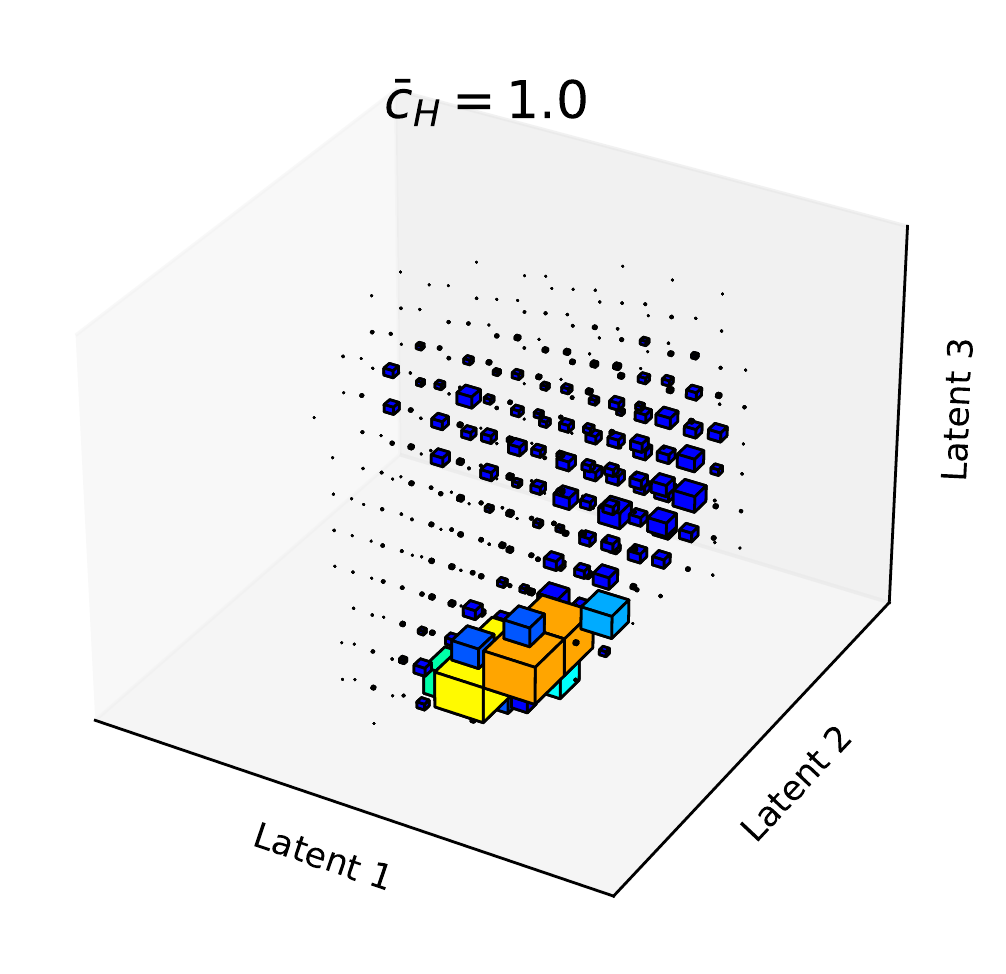}
\caption{ The normalized distributions of the latent features for different processes in the dileptonic channel (upper panels) and semi-leptonic channel (lower panels). Processes from left to right panels correspond to the backgrounds and EW $W^+ W^- jj$ productions in the EFT model with $\bar{c}_H=-1.0,~0,~1.0$, respectively. \label{fig:chfeaturell}}
\end{figure}

In this and the next section, the same network that is trained on the labeled SM background processes as well as the SM $W^{\pm}W^{\mp}jj$ with different polarizations is used for testing. 
Events of the new physics are not used for training the network, in order to show that our method is model agnostic.
Analyzing the preselected events of both SM background processes and the EFT processes with the pre-trained network, we can obtain the distributions of those processes in the 3-dimensional latent space.
The normalized distributions are presented in Fig.~\ref{fig:chfeaturell}, where the background corresponds to the weighted sum of all SM processes (including $t t_\ell$, $t W_\ell$/$t_\ell W$, $W_\ell W j j^{\rm QCD}$, $W_\ell Z j j^{\rm QCD}$ and $W_\ell Z j j^{\rm EW}$) as discussed in Sec.~\ref{sec:events}.
Since the network is trained to classify the SM background processes with the SM $W Wj j^{\rm EW}$, it is not surprised to find that the background events are well separated from the signal events (EW $W Wj j$ production in the EFT).
Moreover, there are visible differences among the distributions of EW $W Wj j$ production with different $\bar{c}_H$. This feature can be used to constrain the value of $\bar{c}_H$.

\begin{figure}[htbp]
\centering
\includegraphics[height=5cm,width=6cm]{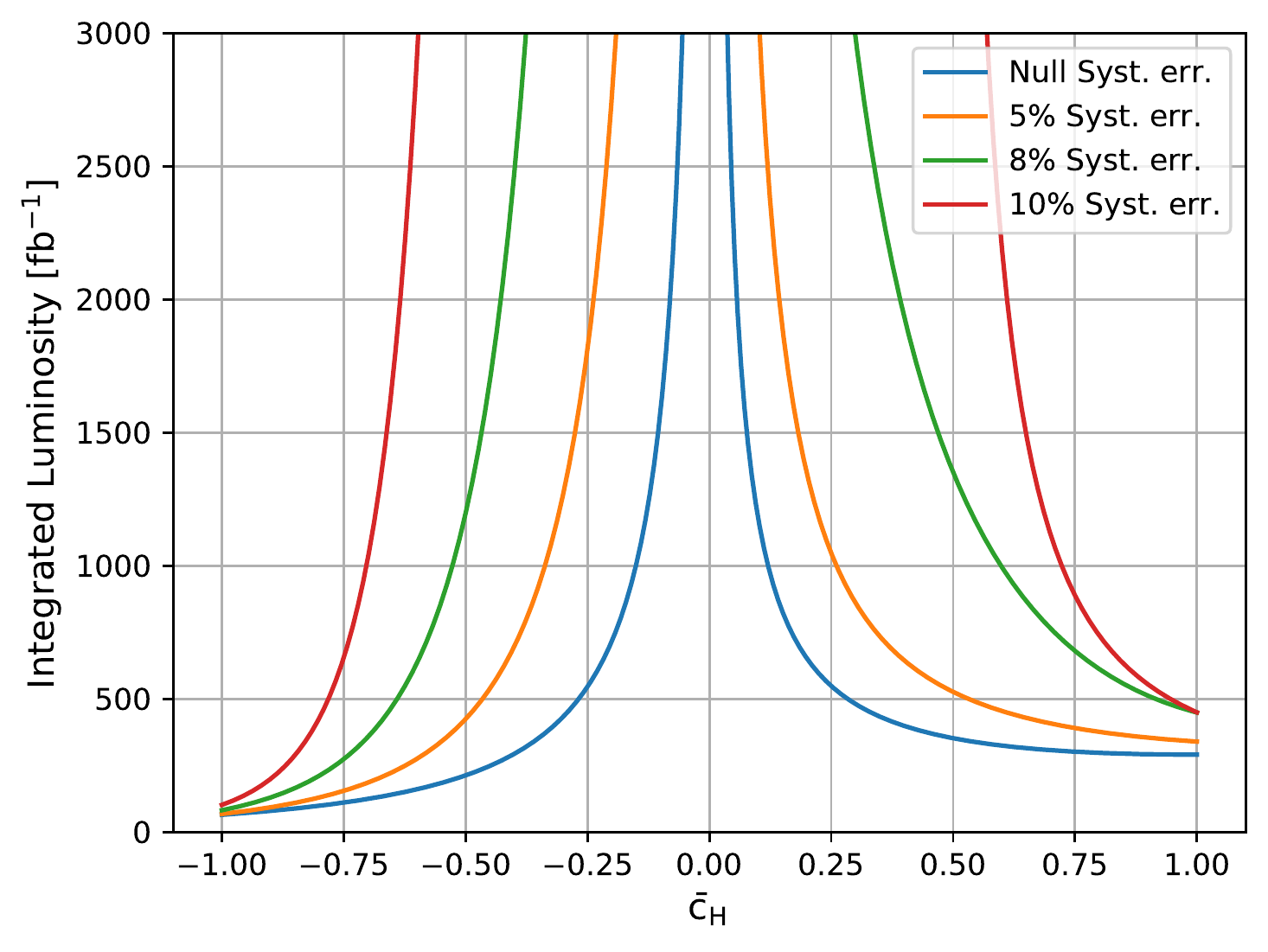}
\includegraphics[height=5cm,width=6cm]{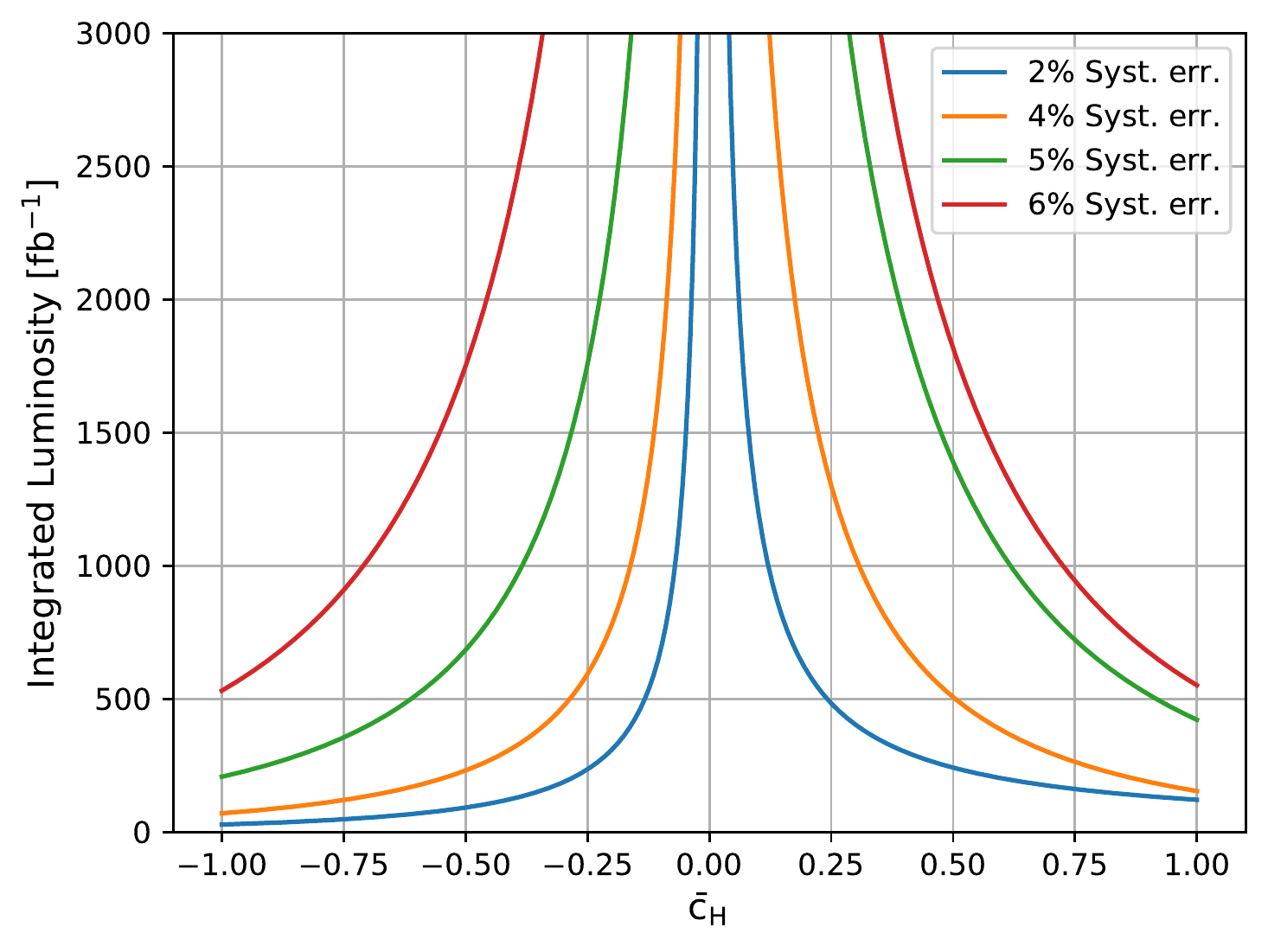}
\caption{\label{fig:chpvalue} Integrated luminosity required to probe the signal (with different $\bar{c}_H$) at 95\% C.L. in the dileptonic channel (left panel) and semi-leptonic channel (right panel). Several different systematic uncertainties are considered. }
\end{figure}

To measure the consistency of the SM and EFT with non-zero $\bar{c}_H$, we perform the binned log-likelihood test in the latent space. As have been discussed in Sec.~\ref{sec:likelihood}, only ten bins with highest signal to background ratios are used. According to our simulation, this will select $\sim$ 30\% signal events and $\sim$ 0.5\% background events after the preselection.
The null hypothesis is the SM backgrounds plus SM EW $W^+W^- j j$ and the test hypothesis is the SM backgrounds plus EFT EW $W^+W^- j j$ with a non-zero $\bar{c}_H$.
The required integrated luminosity to achieve 95\% Confidence Level (C.L.) probing for different $\bar{c}_H$ are presented in Fig.~\ref{fig:chpvalue}.
It can be seen that the semi-leptonic channel outperforms the dileptonic channel if the systematic uncertainty can be controlled below $\sim$ 5\%. Due to higher backgrounds in the semi-leptonic channel, the sensitivity drop quickly when the systematic uncertainty is larger than 5\%.
With systematic uncertainty around 5\%, our method will be able to constrain the $\bar{c}_H$ to [-0.2,0.1] at high luminosity LHC.

\subsection{Effects of event simulation error}
\begin{figure}[htb]
\centering
\includegraphics[width=0.45\textwidth]{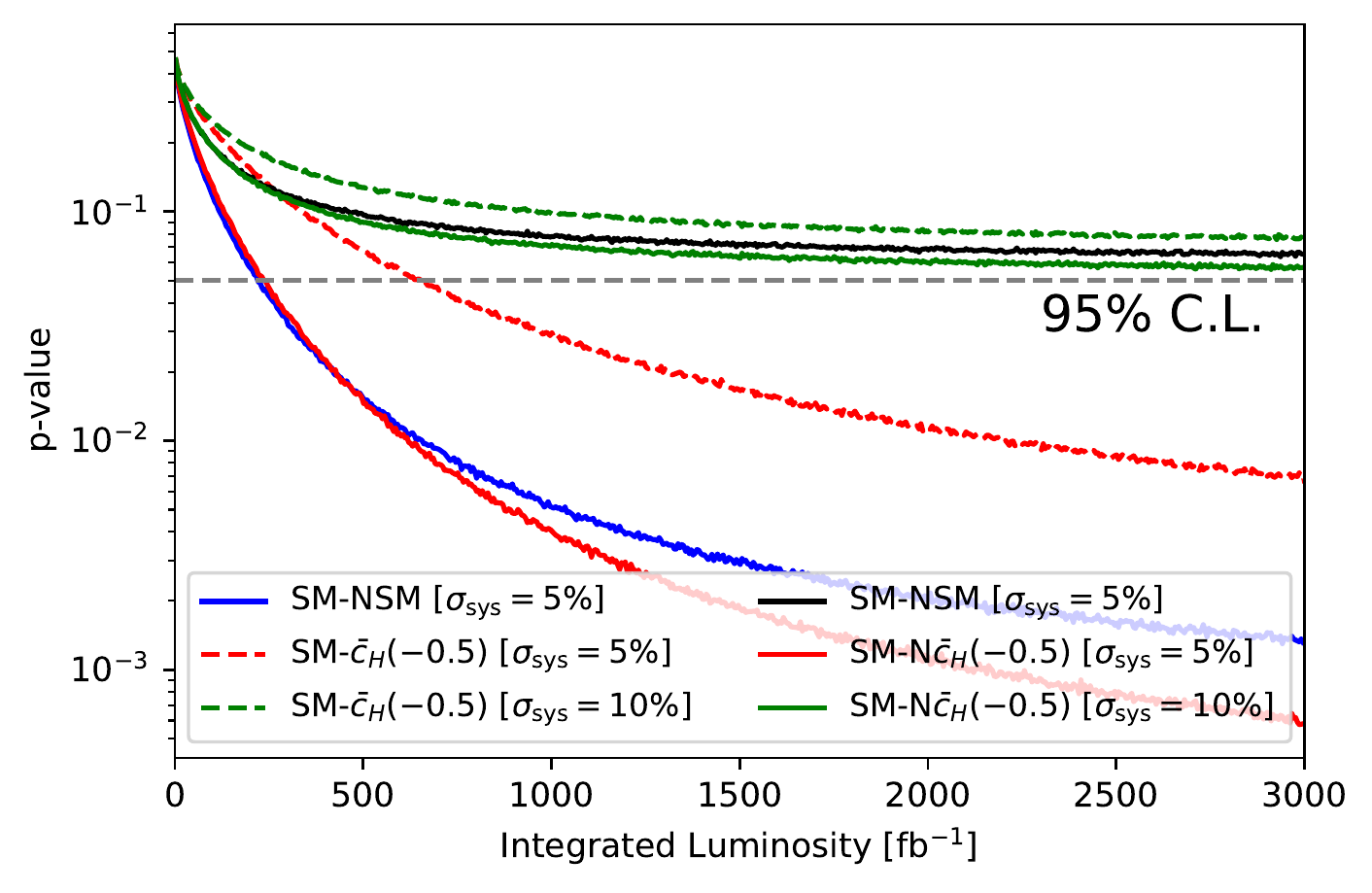}
\includegraphics[width=0.45\textwidth]{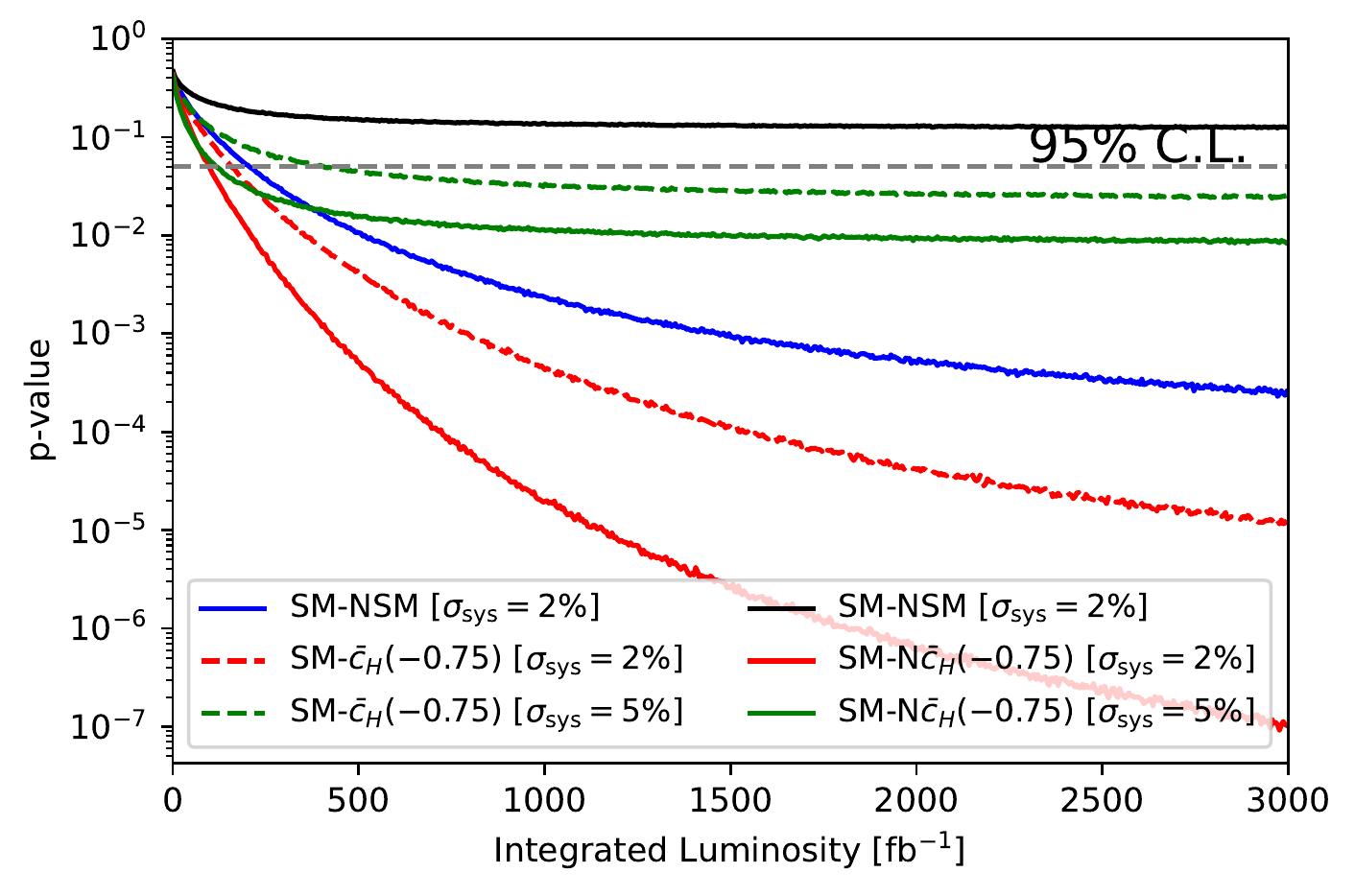}
\caption{\label{fig:nbg} The $p$-values (at varying integrated luminosity) for independent simulation of event samples and for benchmark points. 
Left panel: dileptonic channel with benchmark point $\bar{c}_H =-0.5$. Right panel: semileptonic channel with benchmark point $\bar{c}_H =-0.75$. 
SM denotes the null hypothesis with events simulated as before. For NSM and N$\bar{c}_H$, the background events are simulated by using Herwig++. 
The $\sigma_{\text{sys}}$ indicates the systematic uncertainty that we adopt in the binned log-likelihood analysis. }
\end{figure}

Since our network is trained to detect the anomaly in the simulated SM processes, it could be sensitive to the errors in the simulation. In Fig.~\ref{fig:nbg}, we show how the results of our shape analyses change if the testing samples are simulated independently from the training ones. 
To calculate the $p$-values in the figure, the null hypothesis is always the SM prediction with events simulations as have been discussed above. 
In test hypothesis (NSM and N$\bar{c}_H$), the events of the SM processes are simulated independently with Herwig++~\cite{Bahr:2008pv,Bellm:2015jjp} for parton shower and hadronization, and Delphes with ATLAS parameters for detector simulation.  
For the SM processes, two independent simulations lead to $5\%$ ($3\%$) systematical deviations in the selected bins for dileptonic (semileptonic) channel. 
As a result, if the systematic uncertainty in the shape analysis is chosen to be smaller than the systematical deviations caused by the simulation, event samples of two simulations for the SM processes can be distinguished, as shown by the blues lines in both panels. 
Moreover, the difference of the simulations in null and test hypothesis renders over-optimistic results for the sensitivity to new physics, although the effects is mild when the systematic uncertainty in the shape analysis is chosen to be large.

\section{Application to the 2HDM} \label{sec:2hdm}

The EFT description may not valid when the collision energy is approaching the masses of new states. Here we consider an ultraviolet complete model, the 2HDM~\cite{Aoki:2009ha,Branco:2011iw} which is one of the simplest extension to the Higgs sector of the SM.
The scalar sector of the 2HDM consists of two $SU_{W}(2)$ doublets.
A discrete $Z_2$ symmetry is imposed to avoid tree-level flavor changing neutral currents. Depending on how this symmetry is extended to the fermion sector, four types of the 2HDM can be realized. The type-II case will be considered in this work.
The 2HDM predicts many remarkable signatures at the hadron collider. In particular, there are resonant signals due to the existence of extra CP-even scalar, CP-odd scalar and charged scalar.
Instead of proposing dedicated search for each of those signals, we will show that our method is sensitive to changes of the polarization and kinematic properties of the EW $W^+ W^- j j$ production in the 2HDM.
Comparing the latent features of the $W^+ W^- j j$ process in the 2HDM with those from measurement, constraints on the parameters of the 2HDM can be obtained.

There are six parameters in the type-II 2HDM: mass of scalars ($m_{H_1}, m_{H_2}$, $m_A$ and $m_{H^\pm}$), the mixing angle between two CP-even scalars $\alpha$ and the ratio between two vacuum expectation values $\tan \beta$.
The $m_{H_1}$ has been measured to be close to 125 GeV. The $m_A$ and $m_{H^\pm}$ are not relevant in the $W^+ W^- j j$ production. Their mass is set to 3 TeV to forbid the decays of $H_2$ into those states.
The couplings of CP-even scalars to the $W$ bosons are given by
\begin{align}
\mathcal{L} \supset  \frac{2 m^2_W}{v} \sin (\alpha-\beta) H_1 W^{+}_{\mu} W^{\mu -} + \frac{2 m^2_W}{v} \cos (\alpha-\beta) H_2 W^{+}_{\mu} W^{\mu -}~.
\end{align}
So the combination $\sin (\alpha-\beta)$ is usually used to replace the $\alpha$ parameter.
Even though the $\tan \beta$ alone is not related to the $HWW$ couplings, it can modify the scalar to fermions couplings,  which means the total decay width of the $H_2$ thus the kinematics of $W^+ W^- jj$ can be affected.
We will chose $\tan \beta=5$ for simplicity~\footnote{The influence of the $\tan \beta$ to the $W^+ W^- jj$ production is mild as long as the decay width of the $H_2$ is not too large.}. So we are left with two free parameters: $m_{H_2}$ and $\sin (\alpha-\beta)$. The partial widths of the $H_2$ are given by
\begin{align}
\Gamma(H_2 \to WW) &= \frac{g_w^4  \cos^2(\alpha-\beta) v^2 }{256  \pi m_{H_2}} \sqrt{1- 4 \frac{m_W^2}{m^2_{H_2}}} \frac{m_{H_2}^4 - 4 m_{H_2}^2 m^2_W +12 m^4_W }{m^4_W} ~,\\
\Gamma(H_2 \to ZZ) &= \frac{(g')^4 \cos^2(\alpha-\beta) v^2}{512 \pi m_{H_2}}  \sqrt{1-4 \frac{m^2_Z}{m^2_{H_2}}} \frac{m_{H_2}^4-4m_{H_2}^2 m_Z^2 +12 m^4_Z}{m^4_Z} ~,\\
\Gamma(H_2 \to t\bar{t}) &= \frac{3 y_t^2 (\sin(\beta - \alpha)- \tan \beta \cos(\beta-\alpha))^2}{16 \pi m_{H_2} \tan^2 \beta} \sqrt{1-\frac{4 m_t^2}{m^2_{H_2}}} (m^2_{H_2} - 4 m^2_t) ~, \\
\Gamma(H_2 \to b\bar{b}) &= \frac{3 y_b^2 (\cos(\beta - \alpha)- \tan \beta \sin(\beta-\alpha))^2}{16 \pi m_{H_2} } \sqrt{1-\frac{4 m_b^2}{m^2_{H_2}}} (m^2_{H_2} - 4 m^2_b) ~,
\end{align}
with $g'=\cos(\theta_w) g_w + \sin(\theta_w) g_1$, and $y_t$/$y_b$ is the Yukawa coupling of the top/bottom quark.

The model is implemented in \textsc{FeynRules}~\cite{Alloul:2013bka}, which generates the UFO model files for the \textsc{MG5} to calculate the leading order production cross section and simulate the events. In Tab.~\ref{2hdm_xsec}, we present the production cross sections of the EW $W^+ W^- jj$ process for a few points in the 2HDM as illustration. In particular, the contribution of the heavy scalar $H_2$ is taken into account, which lead to an increased total production rate for the most of the time~\footnote{The cross section in 2HDM can be smaller than that in SM when the mass of the $H_2$ is heavy and decay width of the $H_2$ is large, because of the destructive interference between $H_1$ and $H_2$ in some phase space.}.

\begin{table}[htb]
\begin{center}
\begin{tabular}{|c||c|c|c|c|c|c|} \hline
$(m_{h_2}, \sin(\beta - \alpha))$ &  (300,0.7) & (300,0.9) & (700, 0.7) & (700,0.9) \\ \hline \hline
$\sigma^0_{m_{jj}>500}$ [fb] & 636.2 & 492.5 & 461.9 & 428.5  \\ \hline
$\sigma_{ll}$ [fb] & 8.362 & 5.853 &  5.527 & 4.842 \\ \hline
$\sigma_{lj}$ [fb]  & 64.07 & 46.52 & 43.70 &  39.33 \\ \hline \hline
$\sigma^{LL}_{m_{jj}>500}$ [fb] & 170.75 & 79.81 & 71.58 & 42.65 \\ \hline
$\sigma^{LL}_{ll}$ [fb] & 2.91 &  1.27 &  1.30 &  0.676 \\ \hline
$\sigma^{LL}_{lj}$ [fb] & 20.78  & 9.35 &  9.50 &  5.06 \\ \hline
\end{tabular}
\caption{\label{2hdm_xsec}  Similar as Tab.~\ref{tab:ch_xsec}, but for the 2HDM model. The corresponding parameters are given in the first row. }
\end{center}
\end{table}

Due to the facts that the cancellation between the amplitudes with and without Higgs exchange are delayed to the scale of $m_{H_2}$ and the heavy scalar dominantly decays into longitudinally polarized vector boson, the fraction of $W^+_L W^-_L jj$ is considerably larger than that of the SM.
For relatively light $H_2$ and small $\sin(\beta-\alpha)$ (which means the contribution of $H_2$ is significant), the fraction of $W^+_L W^-_L jj$ can reach $\sim$ 30\% before applying the preselection cuts, while the number is 6\% in the SM.
The preselections can increase the fraction even further.
This feature renders our network very sensitive to the signals in the 2HDM.

\begin{figure}[htbp]
\centering
\includegraphics[width=0.32\textwidth]{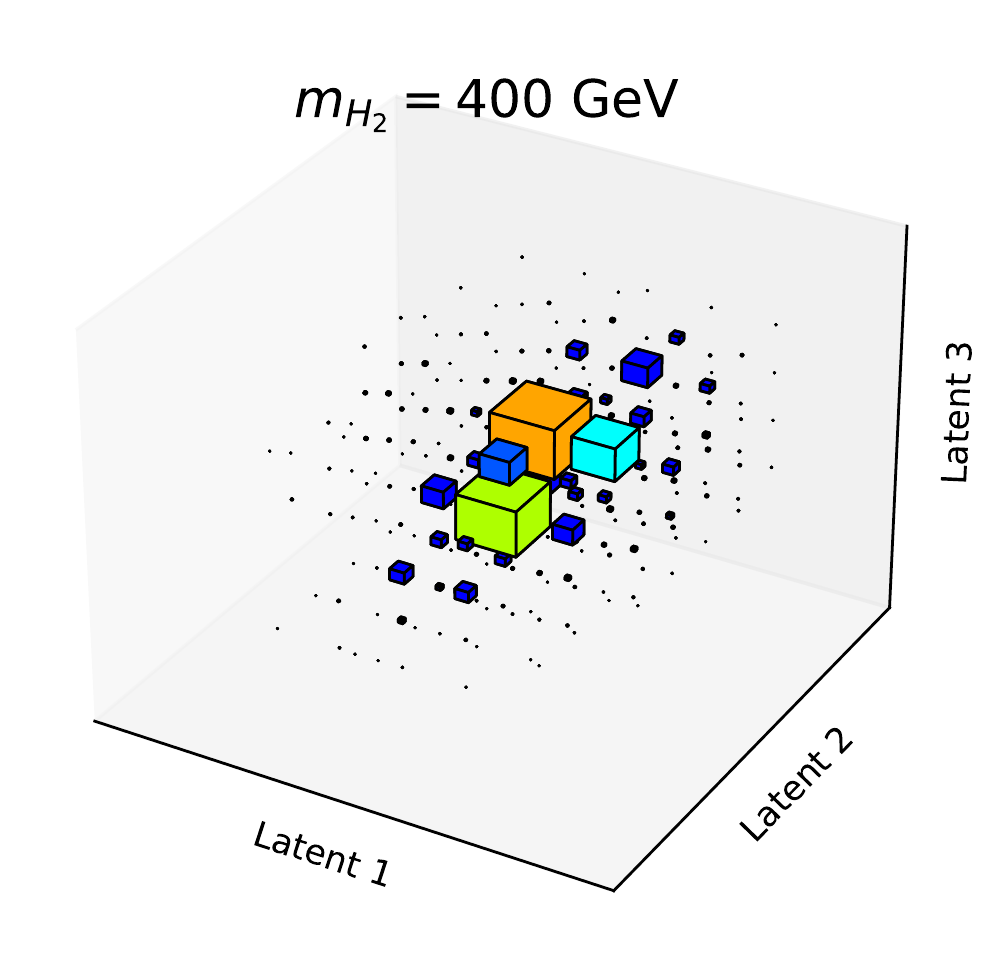}
\includegraphics[width=0.32\textwidth]{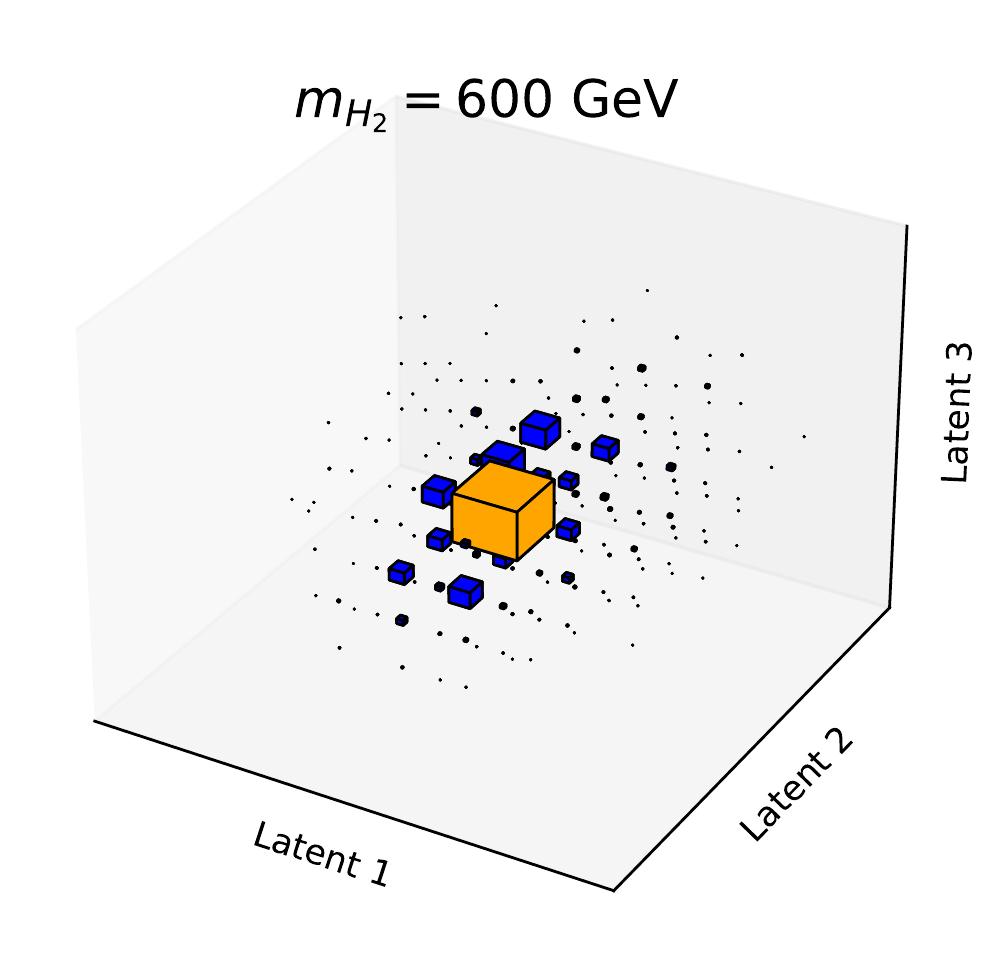}
\includegraphics[width=0.32\textwidth]{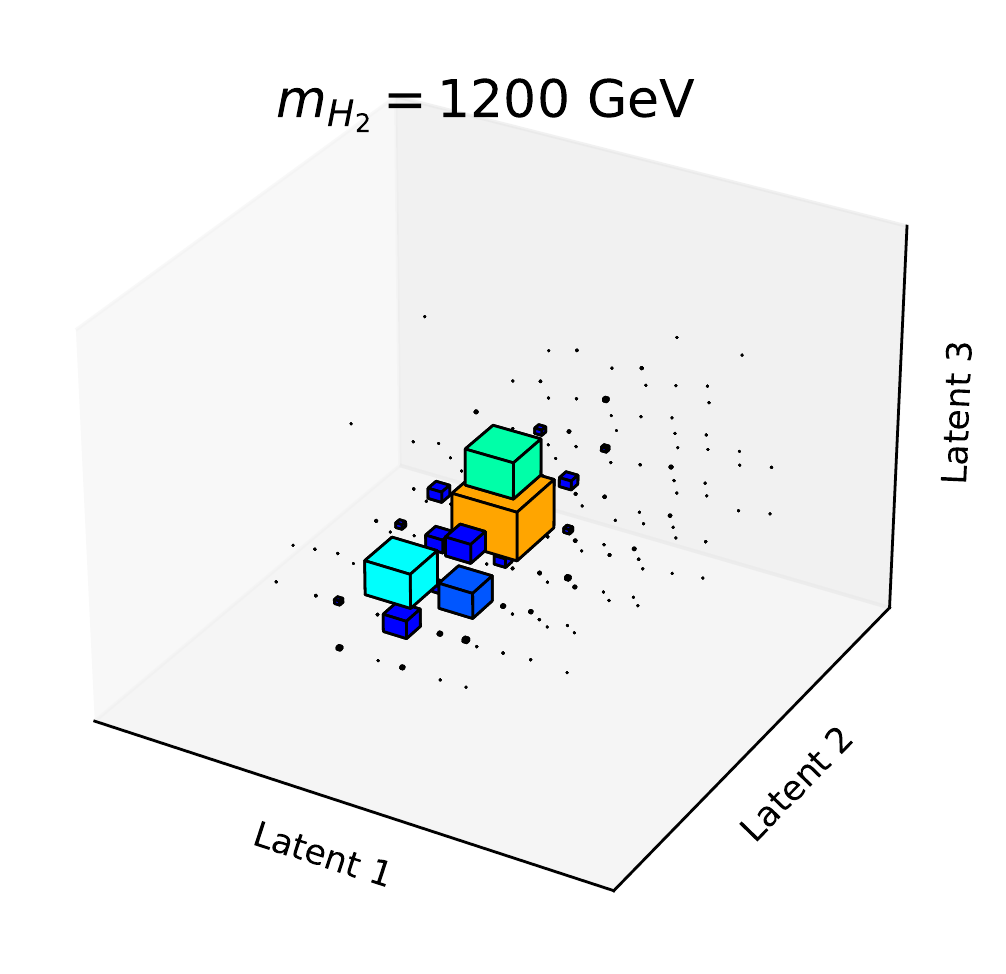}
\caption{\label{fig:peak} The normalized distributions of the latent features for the resonant $H_2$ production and decay $H_2 \to W^+ W^-$ in the dileptonic channel. The mass of the $H_2$ is given in the title  of each subfigure. }
\end{figure}

Moreover, the existence of the $H_2$ resonance in the $W^+W^- jj$ production also gives rise to discriminative features in the final state.
In Fig.~\ref{fig:peak}, we plot the normalized distributions of latent features for the $W^+W^- jj$ production from pure $H_2$ resonance in the dileptonic channel. Different masses of the $H_2$ have distinct distributions in the latent space.
It means the network is not only capable of classifying the polarizations of the vector bosons, but also sensitive to their kinematic properties, even though those 2HDM events are not used for trainning.

\begin{figure}[tbp]
\centering
\includegraphics[height=5cm,width=6cm]{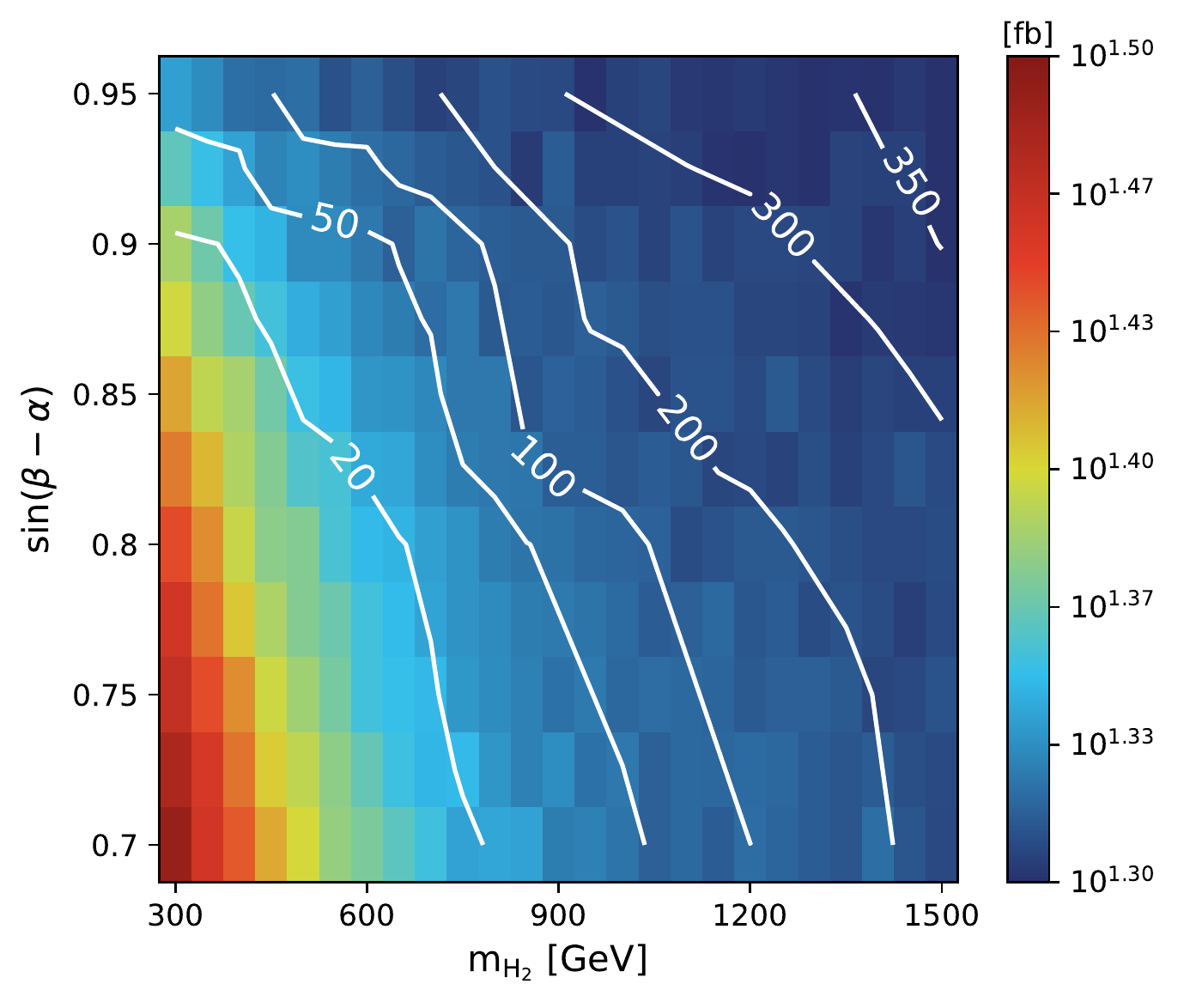}
\includegraphics[height=5cm,width=6cm]{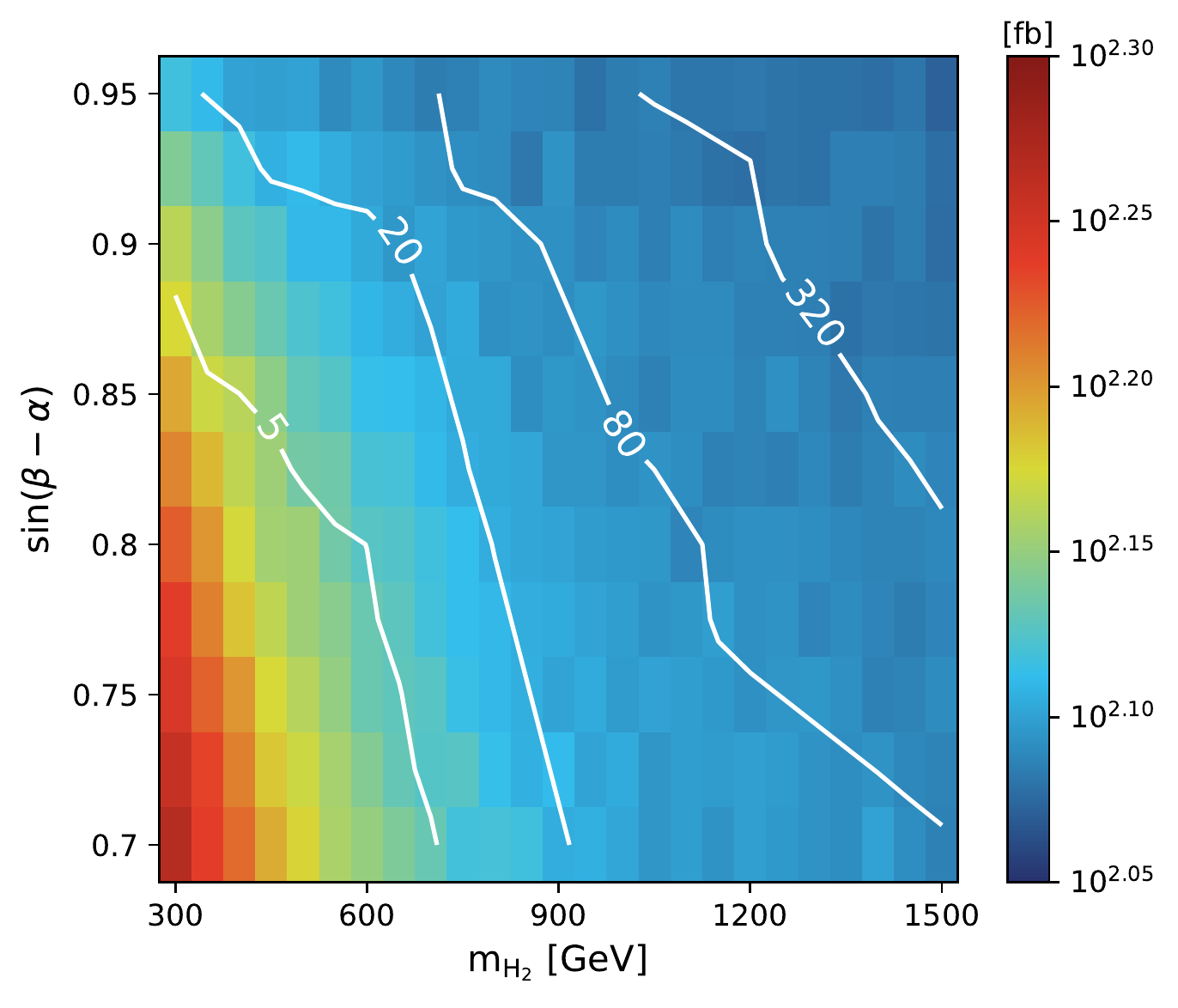}
\caption{\label{fig:2hdmpvalue} The contours correspond to the required integrated luminosity to probe the signal (with different $\sin(\beta - \alpha)$ and $m_{H_2}$) at 95\% C.L. The color grades correspond to the fiducial cross sections (requiring $m_{jj} >500$ GeV at parton level) times the branching ratios. The systematic uncertainties are set to 5\% for both the dileptonic channel (left) and semi-leptonic channel (right).}
\end{figure}

Finally, we pass the preselected events in dileptonic channel and semi-leptonic channel to the pre-trained network, to extract the latent features. The binned log-likelihood test is performed in the latent space to find out the discovery potential of models with different parameters in 2HDM.
Similar as before, the null hypothesis is taken as the SM backgrounds plus the SM EW $W^+ W^- j j$ and the test hypothesis is taken as the SM backgrounds (assuming those processes are kept intact in 2HDM) plus the EW $W^+ W^- j j$ in 2HDM with different sets of parameters.
The required integrated luminosity for achieving 95\% C.L. probing on the $m_{H_2}$-$\sin(\beta-\alpha)$ plane are shown in Fig.~\ref{fig:2hdmpvalue}, for dileptonic channel and semi-leptonic channel, respectively.
Unlike the traditional heavy Higgs resonant searches~\cite{Aaboud:2018bun,Sirunyan:2019pqw}, the sensitivities of which drop quickly at large $m_{H_2}$ due to the suppressed production rate. Our method probe both the resonant feature and the modification to Higgs couplings simultaneously. The parameter space with $H_2$ as heavy as 1.5 TeV can be probed with relatively low integrated luminosity provided the $\sin(\beta-\alpha)$ is not too close to one.
However, as $\sin(\beta-\alpha) \to 1$ (the alignment limit), our method loss the sensitivity completely. 
Searches for the resonances in fermionic channels are still able to constrain the model~\cite{CMS:2019hvr, Aad:2020zxo,Kling:2020hmi,Chen:2015fca}, since their productions are mainly controlled by the Yukawa couplings.
The production cross sections of both channel before applying the preselection cuts are indicated by the color grades in the figure. We can find the sensitivity of the method is roughly determined by the cross section, even though a slightly better sensitivity can be achieved in the small $\sin(\beta - \alpha)$ region, e.g. comparing to the the point ($m_{H_2}=300~\text{GeV}, \sin (\beta -\alpha)=0.9$),  lower integrated luminosity is required to probe the point ($m_{H_2}=550~\text{GeV}, \sin (\beta -\alpha)=0.7$), even though their production cross sections are similar. The improvement of the sensitivity attribute to the fact that point with smaller $\sin (\beta -\alpha)=0.7$ contains larger fraction of the longitudinal $W$ boson.

\section{Discussion and conclusion} \label{sec:conclude}
\begin{figure}[htb]
\centering
\includegraphics[width=0.45\textwidth]{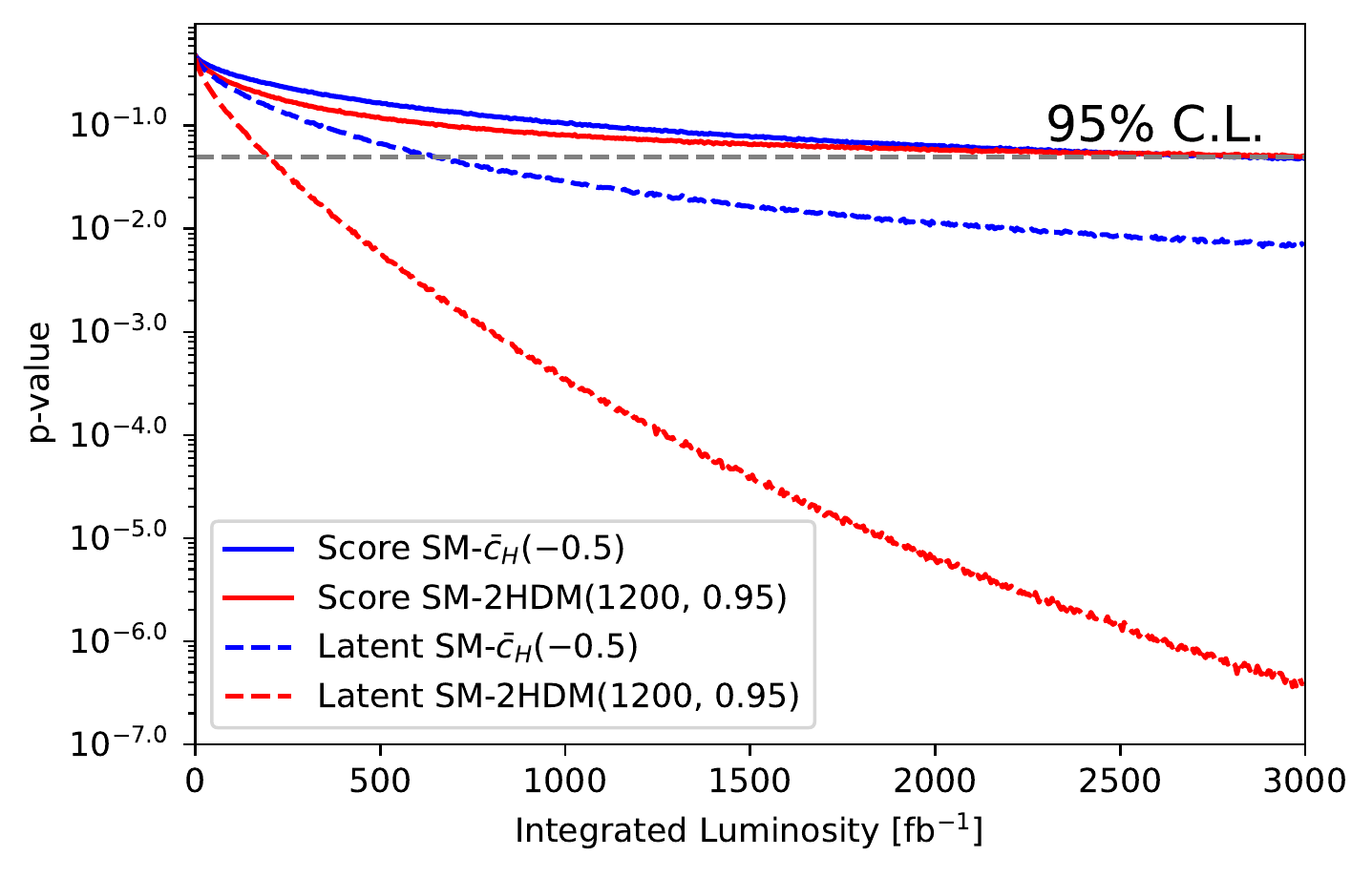}
\includegraphics[width=0.45\textwidth]{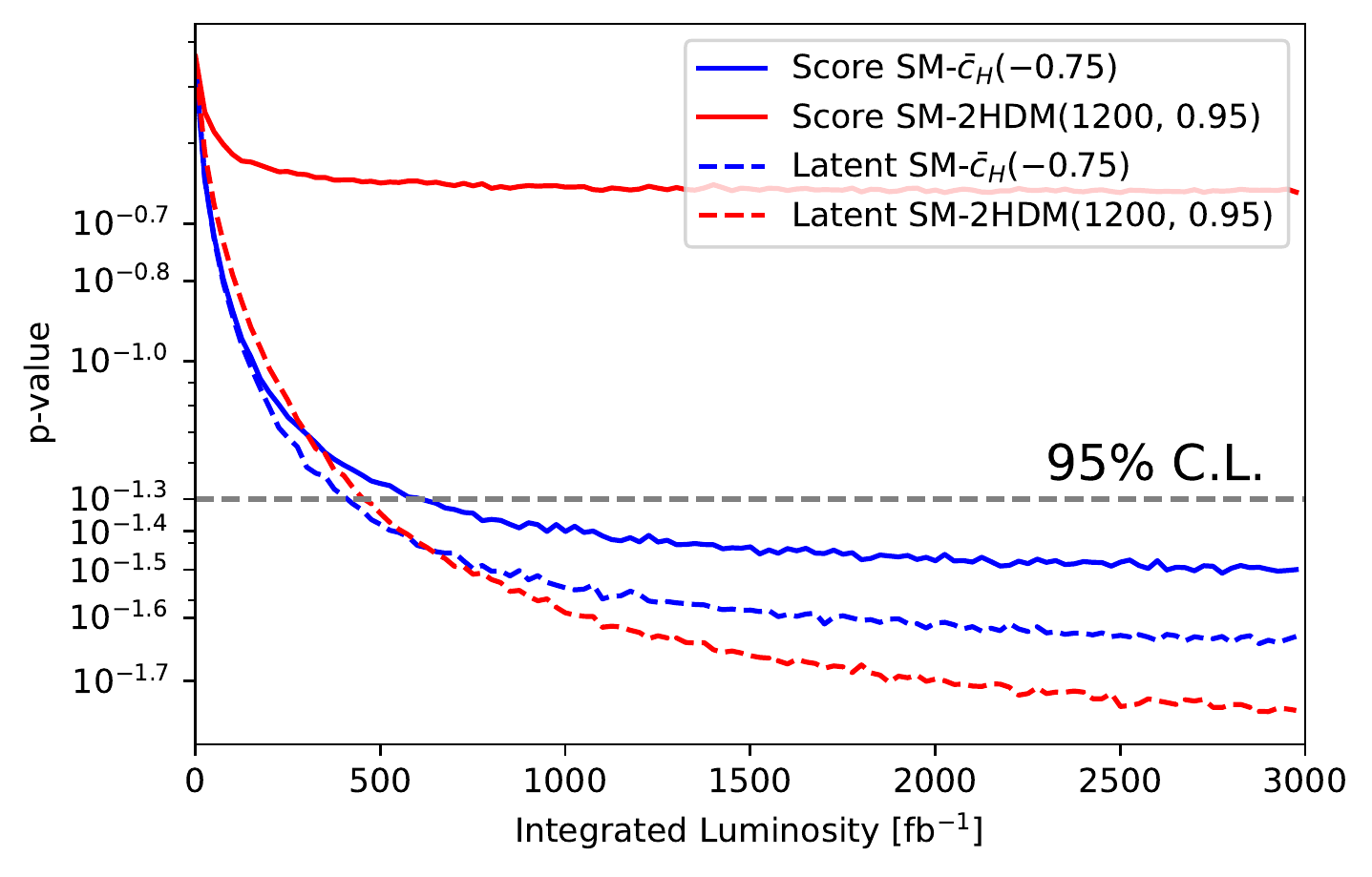}
\caption{\label{fig:pscore} The $p$-value (at varying integrated luminosity) of the shape analysis in latent space (dashed lines) and the $p$-value calculated only with score from the classifier (solid lines) for dilepton channel (left panel) and semileptonic channel (right panel). }
\end{figure}

In this work, we construct a neural network consists of a classification network and an autoencoder.
With the input of low level information (4-momenta and the identities of particles in our case), the network is capable of reducing the dimensionality of the feature space for $WWjj$ production, without losing much discriminating power (discriminating the EW $WWjj$ from other processes, as well as discriminating different polarization modes of the EW $WWjj$).
We find the feature space of both dileptonic and semi-leptonic channels can be compacted into three dimensions.
Performing the binned log-likelihood test on the distributions of latent features, we can draw the conclusion whether the data is consistent with the SM predict. We have shown that those latent features are very sensitive to various possible new physics contributing to the VBS.
Even though the scores given by the classifier network contain a certain amount of the process information, they are not as complete as the latent features. In Fig.~\ref{fig:pscore}, we present the sensitivities of the latent features and the sensitivities of the score~\footnote{Among the scores, we find the summation of scores of all polarization components of EW $WWjj$ lead to the best result. So it is used for calculating the $p$-value in the plots.} obtained by the classifier for two benchmark points in the EFT and the 2HDM. It is not surprised to find out that the latent features have better sensitivities. In particular, the remarkable kinematic feature of the 2HDM is not very useful in classifying SM processes, which means this sort of information can be lost in the scores given by the classifier. Comparing to the EFT case, the improvements of using latent features are much more significant in the 2HDM model.

Considering both the dileptonic and semi-leptonic channel of the $W^+W^- jj$ production, we show that our network is capable of classifying different polarization modes efficiently.  Without considering the background, the LHC dataset with integrated luminosity $\lesssim 600$ fb$^{-1}$ will be sufficient to probe the 1\% change in the longitudinal $W^+W^- j j $ fraction, using the semi-leptonic channel. The dileptonic channel is less sensitive due to its small production rate.
Then, the network is applied to the EFT with non-zero $\mathcal{O}_H$ operator and the type-II 2HDM taking into account the background effects, to obtain more complete and realistic results.
In the EFT, our method will be able to constrain the coefficient $\bar{c}_H$ to [-0.2,0.1] providing the systematic uncertainty is around 5\%.  The dileptonic channel outperforms the semi-leptonic channel if the systematic uncertainty is higher than 5\%.
In the 2HDM, since our method is sensitive to both the resonant decay $H_2 \to W^+ W^-$ and the modification to the SM Higgs couplings, the whole region with $\sin(\beta - \alpha) \lesssim 0.95$ and $m_{H_2} \lesssim 1.5$ TeV can be probed with integrated luminosity $\sim$ 300 fb$^{-1}$ at the LHC.

We note that modifications of the SM are unlikely to be confined to VBS processes.
Assuming a new physics scenario of some kind, the model dependent searches can be very effective in discovering the signal.
Our method may not as sensitive as those model dependent searches for specific signals. 
For example, in the 2HDM with $\tan\beta=5$, our method is insensitive to the parameter space where $\cos(\beta - \alpha) = 0.05$ (corresponds to $\sin(\beta - \alpha) = 0.9987$). On the other hand, searches for $H\to \tau \tau$ at the LHC have already excluded the parameter space with $m_{H} \sim [200, 350]$ GeV~\cite{CMS:2019hvr, Aad:2020zxo,Kling:2020hmi}.
The advantage of our method is that it is suitable for detecting a wide class of new physics which contributes the VBS, i.e. related to the SM electroweak symmetry breaking.
This is especially useful when the forms of new physics are not known.

\section*{Acknowledgement}
This work was supported in part by the Fundamental Research Funds for the Central Universities, by the NSFC under grant No. 11905149 and No. 11875306. 

\phantomsection
\addcontentsline{toc}{section}{References}
\bibliographystyle{jhep}
\bibliography{ewsb}

\end{document}